\documentclass[twocolumn,showpacs,preprintnumbers,amsmath,amssymb]{revtex4}
                                                                                
\usepackage{graphicx}
\usepackage{dcolumn}
\usepackage{bm}
\usepackage{epsfig}                                                                                

\usepackage{amsmath}
\usepackage{amsfonts}
\usepackage{amssymb}
\usepackage{psfrag}
\usepackage{fancyhdr}

\newcommand\diag{\mbox{diag}} 

\begin{document}

                                                                                
                                                                                


\title{Quasi-2D  perturbations in  duct flows  under transverse magnetic field}
\author{A.Poth\'{e}rat}
\affiliation{
Ilmenau Technical University, Kirchhoffstr. 1 98693 Ilmenau, Germany
}
\email{alban.potherat@tu-ilmenau.de}
\date{8 May 2007}



\begin{abstract}
Inspired by the experiment from \cite{moresco04}, we study the stability of a 
flow of liquid metal in a rectangular, electrically insulating  duct  with a 
steady homogeneous
 magnetic field perpendicular to two of the walls. In this configuration, the 
Lorentz force tends to eliminate the velocity variations in the direction of 
 the magnetic field. This leads to a quasi-two dimensional base flow with 
Hartmann boundary layers near the walls perpendicular to the magnetic field, and
 so-called Shercliff layers in the vicinity of the walls parallel to the field.
Also, the Lorentz force tends to strongly opposes the growth of perturbations with a 
dependence along the magnetic field direction. On these grounds, we represent 
the flow using the model from \cite{sm82}, which essentially consists of 
two-dimensional motion equations with a linear friction term accounting for the 
effect of the Hartmann layer.\\
 The simplicity
of this quasi-2D model makes it possible to study the stability and transient 
growth of quasi-two dimensional perturbations over an extensive range of 
non-dimensional parameters and reach the limit of high magnetic fields. In this 
asymptotic case,  the  Reynolds
 number based on the Shercliff layer thickness $Re/H^{1/2}$ becomes the only 
relevant 
parameter. Tollmien-Schlichting waves are the most linearly unstable mode as for 
the Poiseuille flow, but for $H \gtrsim 42$, a second unstable mode, symmetric 
about the duct axis, appears with a lower growthrate. We find that these layers 
are linearly unstable for 
$Re/H^{1/2}\gtrsim 48350$ and energetically stable for $Re/H^{1/2}\lesssim 65.32$.
Between these two bounds, some non-modal quasi-two dimensional perturbations  
undergo some significant transient growth (between 2 and 7 times more than 
in the case of a purely 2D Poiseuille flow, and for much more subcritical values
of $Re$). In the limit of a high magnetic field, the maximum gain $G_{max}$ 
associated to this transient growth is found to vary as 
$G_{max} \sim (Re/Re_c)^{2/3}$ and occur at time $t_{Gmax}\sim(Re/Re_c)^{1/3}$ 
for streamwise wavenumbers of the same order of magnitude as the critical 
wavenumber for the linear stability.
\end{abstract}

\maketitle


\section{Introduction}
\label{sec:intro}
Liquid metal flows in rectangular ducts under imposed magnetic fields are 
important for the metallurgy as well as for the design of the future ITER nuclear 
fusion reactor. The simplest case of a uniform magnetic field parallel to the 
 side wall of the duct has received considerable attention from theoreticians 
and experimentalists over the last fifty years. \cite{shercliff53} and 
\cite{hunt65} clearly identified the three main regions of the flow in the laminar 
regime for the case of insulating walls perpendicular to the magnetic fields. 
The problem is governed by the Reynolds number $Re$  and the Hartmann number $Ha$,
the square of which represents the ratio between the Lorentz and the viscous 
forces, as well as by the aspect ratio of the duct.
For large values of $Ha$ (typically more than 10), 
Hartmann boundary layers with a simple exponential profile and a thickness of 
$Ha^{-1}$ develop along these walls. These were first 
discovered by \cite{hartmann37} and further demonstrated by \cite{shercliff65}. 
Far from 
the walls, the Lorentz force strongly damps the velocity variations along the 
magnetic field lines so that the flow is two-dimensional, as explained by 
\cite{sm82} and \cite{dav97}. The boundary layers which arise along the 
side walls, now called Shercliff layers, have a complex three-dimensional 
profile 
or thickness $a/Ha^{-1/2}$ ($a$ is the duct dimension along the field) found 
analytically by \cite{shercliff53}.\\
The next step was to investigate the stability of the flow. Probably because 
of its simpler base profile, the Hartmann layer has received the most 
attention: linear stability analyses performed by several authors agree to a 
critical Reynolds number around  $Re/Ha\simeq48000$ 
\cite{lock55,takashima96,lingwood99}, 
whereas the energy 
stability analysis provides a sufficient condition for stability for 
$Re/Ha<25$ \cite{pavlov72,simkhovich74,lingwood99}. More recently, it has been 
suggested that  
some non-modal perturbations may undergo significant transient growth 
\cite{gerard-varet02, airiau04}, and DNS performed by \cite{krasnov04} show 
that the non linear evolution of such perturbations leads to a destabilisation 
of the Hartmann layer for $Re/Ha>390$. All of these works, however, involve 
either a channel flow geometry or a flow over an infinite plate, none of 
which include any Shercliff layers. This has the advantage of clarifying the 
properties of the Hartmann layer itself but raises difficult questions 
when it comes to comparison with experiments in which the presence 
of side walls, and therefore Shercliff layers, cannot be avoided. Many such
experimental data has been produced on duct flows in the past half century
\cite{murgatroyd53, lykoudis60, branover67}. In particular, the most recent 
experiment from \cite{moresco04} was conducted with the sole purpose of finding
 the instability threshold for the Hartmann layer. It consists of measuring 
the total friction in a toroidal square duct  placed inside a $13T$ 
superconducting magnet. In the limit of large values of $Ha$, they  find the 
instability threshold to be $Re/Ha=380$. The excellent agreement between this 
result 
and the channel flow DNS of \cite{krasnov04} suggests that the stability of the 
Hartmann layer is hardly or not affected by the presence of the side walls. It,
however, leaves open the question of knowing whether the Shercliff layers and the 
core flow are laminar or turbulent when the Hartmann layers destabilise. 
The value of the friction measured in \cite{moresco04} when the Hartmann layer 
is laminar, recovers quite closely the friction given by the laminar Hartmann 
layer theory: this indicates at least that no 3D turbulence exists in 
this regime as this would produce a strong extra Joule and viscous dissipation. 
The possibility, however, that the flow is in a quasi-2D turbulent state 
producing 
only a small extra dissipation cannot be written off. 
This is supported by recent studies performed in Ilmenau on the transient growth of 
perturbations in a channel flow between two parallel walls with a magnetic field
parallel to these walls. It was found that for $Ha \gtrsim 100$, vortices aligned 
 with the magnetic field are the perturbations  undergoing maximum transient 
growth, as opposed to the classical Poiseuille case (without magnetic field) 
where streamwise-independent perturbations are the most amplified (not published yet).\\
In view of these 
considerations, the aim of the present work is to undertake a first step toward 
 tackling the question of the stability of Shercliff layers by studying their
stability to quasi two-dimensional perturbations. By quasi-two dimensional, we 
mean that the velocity field is assumed independent of the coordinate along the 
magnetic field lines (this is usually referred to as the 2D core assumption), 
except in the vicinity of the Hartmann walls (those 
orthogonal to the magnetic field) where it exhibits the classical exponential 
profile of Hartmann layers. To this end, we use the 
model from \cite{sm82} (thereafter SM82) which assumes  quasi-two dimensionality
as well as the fact that the  Hartmann 
layers remain laminar. It provides a two-dimensional equation for the 
flow motion in the average plane orthogonal to the magnetic field where the 
effect of the Hartmann layers is taken into account through a linear friction
 term. Comparisons between theory and experiments have shown that 
 this model renders the two-dimensional dynamics of the parallel layers to a 
very good approximation (in particular, friction and turbulent properties 
\cite{psm00, psm05} as well as some stability properties 
\cite{buhl96,frank01}). 
\cite{psm00} has furthermore demonstrated that the 2D model departs 
from the 3D solution of \cite{hunt65} by less than 10\%. On this basis, 
we use the SM82 model to study  the sensitivity of Shercliff layers to quasi-two 
dimensional perturbations. This means all 
flow patterns derived from this model exhibit a Hartman flow profile in the 
direction of the magnetic field and a velocity field to be determined in the 
plane orthogonal to that direction.\\
The layout of this work is as follows: in section \ref{sec:base}, we briefly 
recall the model from \cite{sm82} and the associated quasi-2D base solution 
for the duct flow. The important question of the relevance of the SM82 model to 
the full 3D profile of the Shercliff layers is also reviewed. We then study 
the linear stability of this solution in 
section \ref{sec:linstab}, which provides a sufficient condition for instability. A necessary condition for instability is obtained from the energy stability 
analysis in section \ref{sec:energy}. Section \ref{sec:transg} is then devoted
to  the search of non-modal perturbations undergoing maximal transient growth. 
We finally come back to the experiment of \cite{moresco04} in conclusion 
and briefly discuss it in the light of this work. 
\section{basic equations: 2D model for the duct flow}
\label{sec:base}
\begin{figure}
\psfrag{Insulating}{$\sigma_W=0$}
\psfrag{B}{$\mathbf B$}
\psfrag{U}{$\mathbf U$}
\psfrag{a}{$a$}
\psfrag{L}{$2L$}
\psfrag{H}{$2H$}
\psfrag{ex}{$\mathbf e_x$}
\psfrag{ey}{$\mathbf e_y$}
\psfrag{ez}{$\mathbf e_z$}
\includegraphics[width=8.5cm]{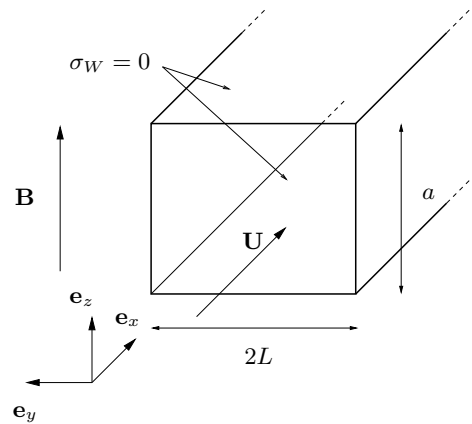}
\caption{Problem geometry: the top and bottom (Hartmann) walls are electrically 
insulating (wall conductivity $\sigma_W=0$ and the electric current normal to the side walls is imposed.}
\label{fig:3dconfig}
\end{figure}

We consider an electrically conducting fluid (electric conductivity $\sigma$, 
kinematic viscosity $\nu$, density $\rho$) flowing through a duct of 
rectangular 
cross section 
(height $a$, width $2L$) and subjected to a 
steady homogeneous magnetic field
perpendicular to the top and bottom walls of the duct (as sketched on figure 
\ref{fig:3dconfig}). 
For a strong enough magnetic field, 
this flow is known to be quasi two-dimensional, that is the velocity 
hardly varies along the direction of the magnetic field, except in Hartmann 
boundary layers which develop along the top and bottom plates. Assuming  
these layers remain laminar, the flow is well described by averaging the 
quasi-static MHD equations along the magnetic field direction. This results in 
the Sommeria and Moreau model (SM82) governing the evolution of the average velocity, 
given here in non-dimensional form  \cite{sm82,psm00}:
\begin{eqnarray}
\nabla \cdot \mathbf u&=&0
\nonumber\\ 
\partial_t \mathbf u + \mathbf u \cdot \nabla \mathbf u + \nabla p
&=&\frac{1}{Re} \left(\nabla^2 \mathbf u-H\mathbf u+ \mathbf f\right)
\label{eq:sm82}
\end{eqnarray}
$Re$ is the Reynolds number $Re=U_0L/\nu$ based on the maximum velocity of the 
base profile $U_0$. The parameter 
$H=n(L^2/a^2) aB\sqrt{\sigma/(\rho\nu)}$ is a measure of the 
 friction term (with characteristic dimensional Hartmann friction 
time $t_H=na/B \sqrt{\rho/(\sigma \nu)}$) which  represents the effect of the 
Lorentz force on the flow. $n$ represents the number of Hartmann layers in the 
problem: here $n=2$ but the case of a rigid upper free surface could be 
handled by setting $n=1$.
The flow is driven by a constant force density $\mathbf f$  (non-dimensional) 
which can result either from a pressure gradient $G$ imposed along the duct 
(\textit{i.e.} dimensionally,  $\mathbf f_{dim}=G \mathbf e_x$) or from a transverse 
electric current density $J_0$ imposed at the side walls (\textit{i.e.} dimensionally, 
$\mathbf f_{dim}= J_0B \mathbf e_x$, as in the experiment from \cite{moresco04}). 
The model (\ref{eq:sm82}) can thus 
describe all cases with 
imposed electric current at the side walls, as well as the case of insulating 
walls for which $J_0=0$. The case of a pressure driven flow where the side walls have a 
finite conductivity studied experimentally by \cite{reed89} requires a slightly 
different model such as that from \cite{buhl96}. The simple form of this 
equation is interesting as it places the problem studied here within the more 
general framework of the 2D flows  with an arbitrary linear friction. On these 
grounds, we shall not restrict ourselves  to the high values of $H$ and 
$H^2/Re$ which correspond to a dominant Lorentz force, and for 
which (\ref{eq:sm82}) is indeed a good approximation of the 3D MHD flow.\\
Using the no slip boundary conditions at the side walls
\begin{equation}
\mathbf u (y=-1)=\mathbf u (y=1)=0,
\label{eq:bc}
\end{equation}
The base flow is found as an exact solution of (\ref{eq:sm82}) and (\ref{eq:bc}) of the form
$\mathbf U= U(y)\mathbf e_x$ with:
\begin{equation}
\frac{H}{f}U(y) =1-\frac{\cosh \sqrt Hy}{\cosh \sqrt H}
\label{eq:base1}
\end{equation}
Since velocities are normalised by the maximum velocity of the base profile 
$U_0$, the latter is related to the driving force by:
\begin{equation}
U_0=\frac{L^2}{\rho \nu}\frac{f_{dim}}{H}\left(1-\frac{1}{\cosh \sqrt H}\right)
\label{eq:fu}
\end{equation}
and (\ref{eq:base1}) can be rewritten:
\begin{equation}
U(y) =\frac{\cosh \sqrt H}{\cosh \sqrt H-1}\left(1-\frac{\cosh \sqrt Hy}{\cosh \sqrt H}\right)
\label{eq:base}
\end{equation}

In the limit $H\rightarrow0$, (\ref{eq:base}) recovers the two-dimensional 
Poiseuille profile whereas for high values of $H$, the profile is almost 
flat, except  in the vicinity of the walls located at $y=1$ and $y=-1$ 
where it exhibits boundary layers of thickness $H^{-1/2}$. 
The full 3D solution also features some  boundary layers of the same thickness 
at this location, which 
are now commonly called Shercliff layers. Their physical mechanism can be 
understood as 
follows: \cite{sm82} has shown that the Lorentz force acts so as a to diffuse  
the momentum of a structure of size $l_\perp$ along the magnetic field lines 
over a length $l_\|$ within a characteristic time:
\begin{equation}
\tau_{2D}=\frac{\rho}{\sigma B^2}\frac{l_\|^2}{l_\perp^2}
\end{equation}
This diffusion results from current loops between plans orthogonal to the 
magnetic field. If $\tau_{2D}$ is shorter than all other timescales, in 
particular those of inertia $\tau_u=l_\perp/U$ and of viscous friction 
$\tau_\nu^\|=l_\|^2/\nu$ and $\tau_\nu^\perp=l_\perp^2/\nu$), then  the 
momentum just outside the Hartmann layer is instantaneously diffused to the 
whole core flow and the flow is quasi two-dimensional. The thickness of the 
parallel layers (including Shercliff layers) is precisely the scale at 
which $\tau_{2D}\sim \tau_\nu^\perp$, as they are determined by the balance 
between the Lorentz force and the viscous friction in planes orthogonal to 
the field. This means that in those layers, viscous friction has had time to 
act on the flow before the momentum outside the Hartmann layers has had time 
to diffuse to the rest of the parallel layer. Since the diffusion is not 
complete, the profile of parallel layers is not 2D, and the Shercliff layers 
result from the balance between 
viscous friction and part of the momentum present outside the Hartmann layer.\\
   In the SM82 model, parallel layers result from a balance between the term 
 representing the friction of the Hartmann layer on the flow and viscous 
friction in planes orthogonal to the field: it is therefore a simplification 
of the dynamics of the Shercliff layers, that assumes that the momentum just 
outside the Hartmann layer still diffuses instantly to the whole parallel 
layer. This 
results in 2D layers, in which viscous friction balances the whole of the 
momentum outside the Hartmann layer. In order to evaluate the loss due to this 
simplification, \cite{psm00} has compared these 2D and 3D 
profiles (see fig.2 p81) and has shown that the 3D profile nowhere departs from 
(\ref{eq:base}) by more than about 10\%. This indicates that in spite of 
the action of viscosity, the quasi two-dimensionality assumption in is only 
slightly violated in the Shercliff layers. Since the physics of the 2D model 
and that of the 3D Shercliff layers are therefore 
close - but not quite identical - the SM82 model is expected to provide some 
relevant indications on the 2D dynamics of the 3D Shercliff layers, even though 
it obviously misses the 3D dynamics. This has 
been found to be the case in many instances where theoretical results derived 
from the 2D model have been compared to experiments: \cite{delannoy99, psm05} 
have performed DNSs of (\ref{eq:sm82}) and a refined version of it, and found
that both the friction and fine turbulent properties of the parallel layer are 
recovered in great detail. Perhaps more importantly for the present study, 
the critical Reynolds number and wavelength for the instability of a free 
parallel layer (which exhibits the same kind of three-dimensionality as the 
Shercliff layers)  measured by \cite{frank01} are in excellent agreement with 
the prediction of \cite{buhl96} based on a variant of (\ref{eq:sm82}) taking 
Hartmann wall conductivity into account.\\
On this basis, it is reasonable to expect (\ref{eq:sm82}) to provide a 
relevant description of the dynamics of quasi-2D perturbations in the 
Shercliff layers. This approach should however not be expected to give the last 
word on the stability of those layers. Instead, it should be considered as a 
toy-model that incorporates most, but not all, of the physics of the full 3D 
problem. Ultimately, the effect of three-dimensionality and 3D perturbations 
will have to be determined by full 3D DNS or a numerical resolution of the 3D 
stability problem. Either of these however 
involve some high computational costs, that preclude any parametric study 
or high values of $H$. Such analyses can be performed with the 2D model, and 
the obtained result can be used to guide future 3D computations performed for 
selected values of $H$ and $Re$.\\
A further advantage of the SM82 approach is that it doesn't reflects 
the physics of the MHD problem only, but also that of any 2D flow with linear 
friction of any origin. The results derived in the forthcoming sections 
 are therefore exact mathematical properties of this class of models.
\section{Linear stability analysis}
\label{sec:linstab}
\subsection{Mathematical formulation}
We shall start the stability analysis by studying the stability of (\ref{eq:base}) 
to infinitesimal perturbations. This provides a sufficient condition for instability. 
To this end, the perturbed velocity profile (and according pressure) is decomposed as:
\begin{equation}
\mathbf u = \mathbf U + \hat{ \mathbf u}(y)\exp(i(kx-\omega t))
\label{eq:normal}
\end{equation}
where it has been taken advantage of the invariance in the $x$-direction to 
write the perturbation as a normal mode. The evolution equation for 
$v(y) = \hat{\mathbf u}(y)\cdot \mathbf e_y$ is obtained by linearisation of 
(\ref{eq:sm82}) around the solution (\ref{eq:base}). After elimination of the 
pressure, this reduces to the following eigenvalue problem:
\begin{eqnarray}
\mathcal L_{OS}v &=& -i\omega \mathcal M v
\label{eq:os}\\
-\mathcal L_{OS}&=& ikU\mathcal M +ikU^{\prime\prime}+\frac{1}{Re}\mathcal M^2
+\frac{H}{ Re}\mathcal M
\nonumber\\
\mathcal M &=& k^2-D^2
\nonumber
\end{eqnarray}
with boundary conditions $v(-1)=v(1)=0$.
$D$ is the differentiation operator with respect to $y$, and the $U^\prime$ denotes
the $y-$ derivative of $U$. The Orr-Sommerfeld operator $\mathcal L_{OS}$ only 
differs from the usual 
Orr-Sommerfeld operator which appears in the linear stability analysis of 
 hydrodynamic parallel flows through the additional friction term so both 
problems can be made formally identical by introducing the frictionless eigenvalue 
$\omega_0=\omega-\frac{H}{i Re}$, as noticed by \cite{thess93}.
\subsection{Numerical procedure}
\label{sec:lsnum}
The eigenvalue problem (\ref{eq:os}), is solved numerically using MATLAB. To this 
end, we use a spectral discretisation based on Tchebychev Polynomials in the $y-$ direction, 
making sure that there are a least 10 collocation points in each of the intervals 
$[-1, -1+H^{-1/2}]$ and $[1-H^{-1/2},1]$, and at least 100 within $[-1,1]$. Our numerical 
resolution follows that presented in detail in \cite{schmid01}, and the MATLAB
routines we use to solve the eigenvalue problem are essentially adapted 
from the routines given there. 
For each chosen value of $H \in [0, 10^4]$, we look for the lowest value of $Re$
such that the maximum eigenvalue $\omega_m(k_m)$ has a positive imaginary part, 
corresponding to the first unstable mode (the corresponding quantities are 
thereafter referred to as \textit{critical}). This is done using a simple dichotomy
method, with a relative precision of $10^{-5}$ on both the critical Reynolds 
number $Re_c$ and the critical wavenumber which achieves the maximum growthrate
$k_c$. We tested our algorithm on the 2D Poiseuille problem ($H=0$) and 
recovered the known values of $Re_c(H=0)=5772,22$ and $k_c(H=0)=1.02$ found 
for instance in \cite{schmid01}. We also calculated the critical Reynolds and 
wavenumber with twice as many collocation points as mentioned above for 
$H=1,10,100$ and $1000$. The difference was below the specified relative 
precision of $10^{-5}$. The same method has been used to determine the critical 
curve in the $(Re,k)$ plane, (\textit{i.e.} the smallest and highest values of 
the unstable wavenumbers at given $Re$).
\subsection{Results}
\begin{figure}
\includegraphics[width=8.5cm]{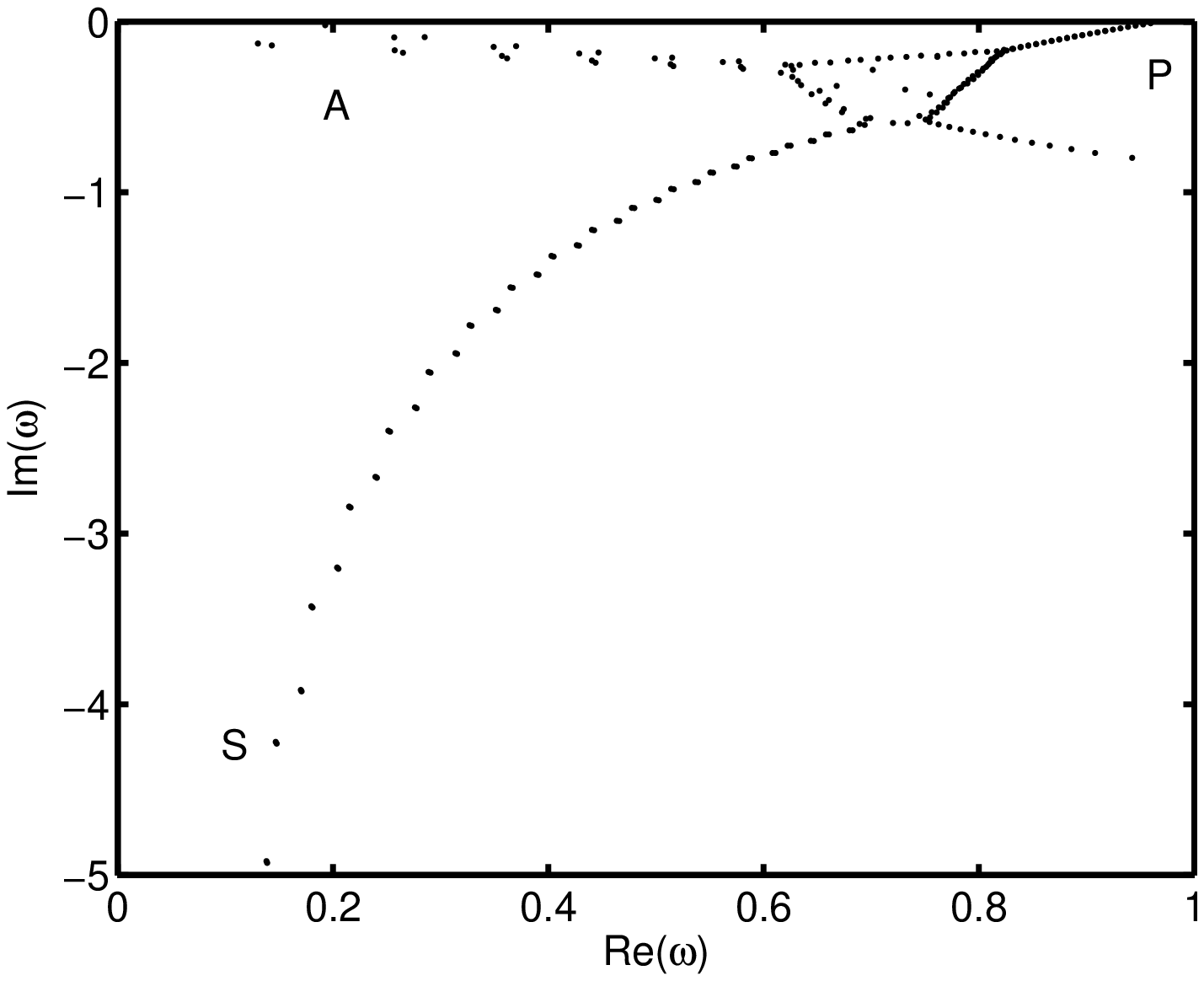}
\includegraphics[width=8.5cm]{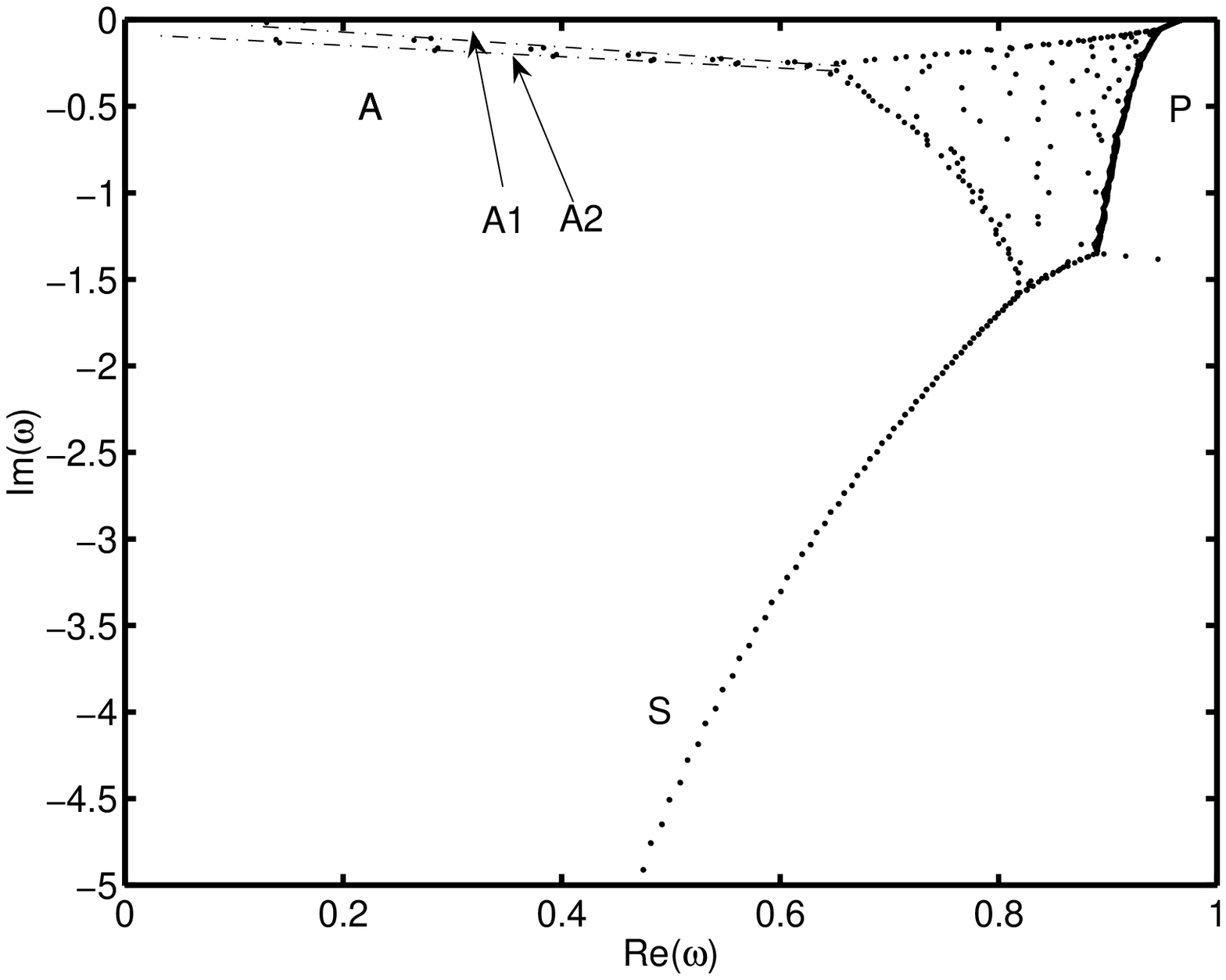}
\caption{Spectra of $\omega$ near criticality for $H=10$ (top) and $H=100$ (bottom). For $H=10$, $Re_c=4.40223. 10^5$ and $k_c=1.73896$. 
}
\label{fig:spectra}
\end{figure}

\begin{figure}
\includegraphics[angle=90,width=3cm,height=12cm]{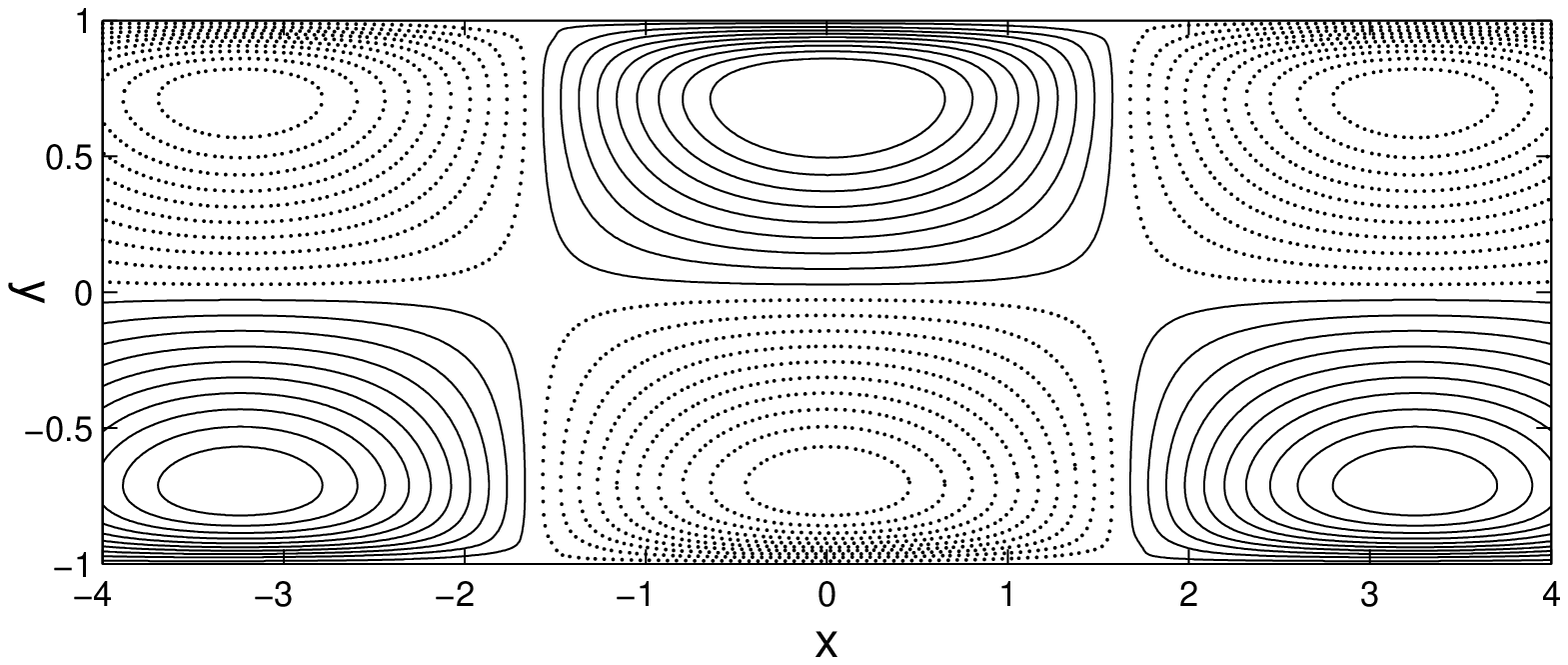}
\includegraphics[angle=90,width=3cm,height=12cm]{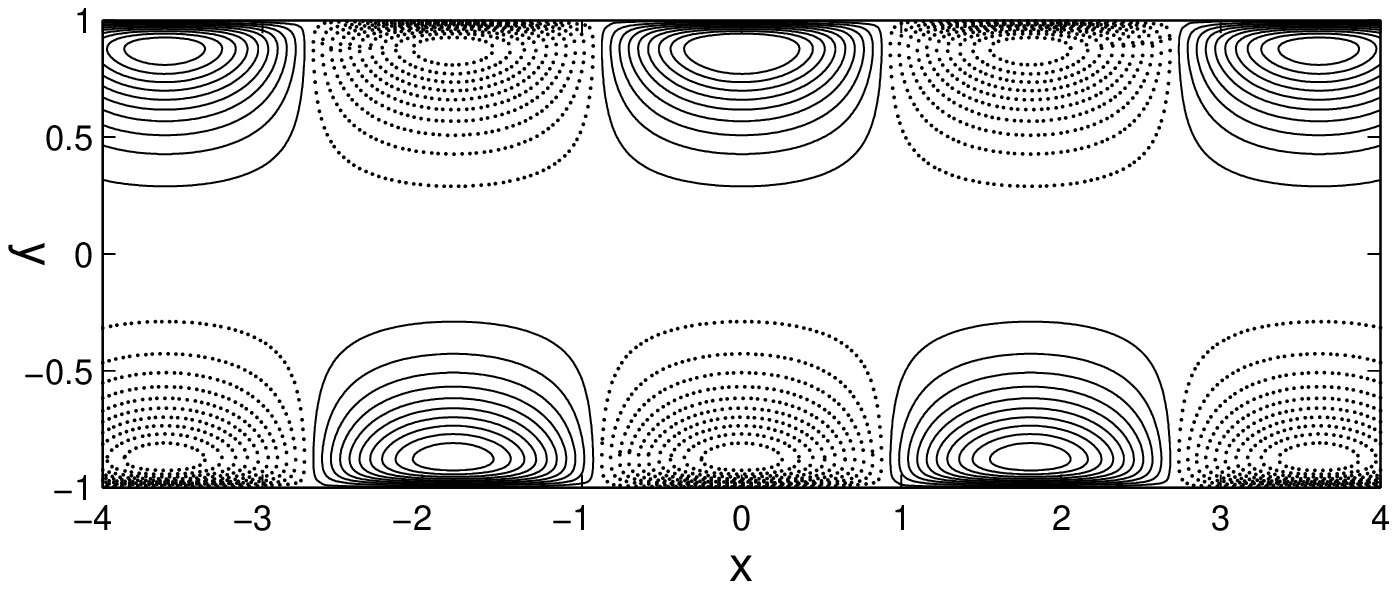}\\
\includegraphics[angle=90,width=3cm,height=12cm]{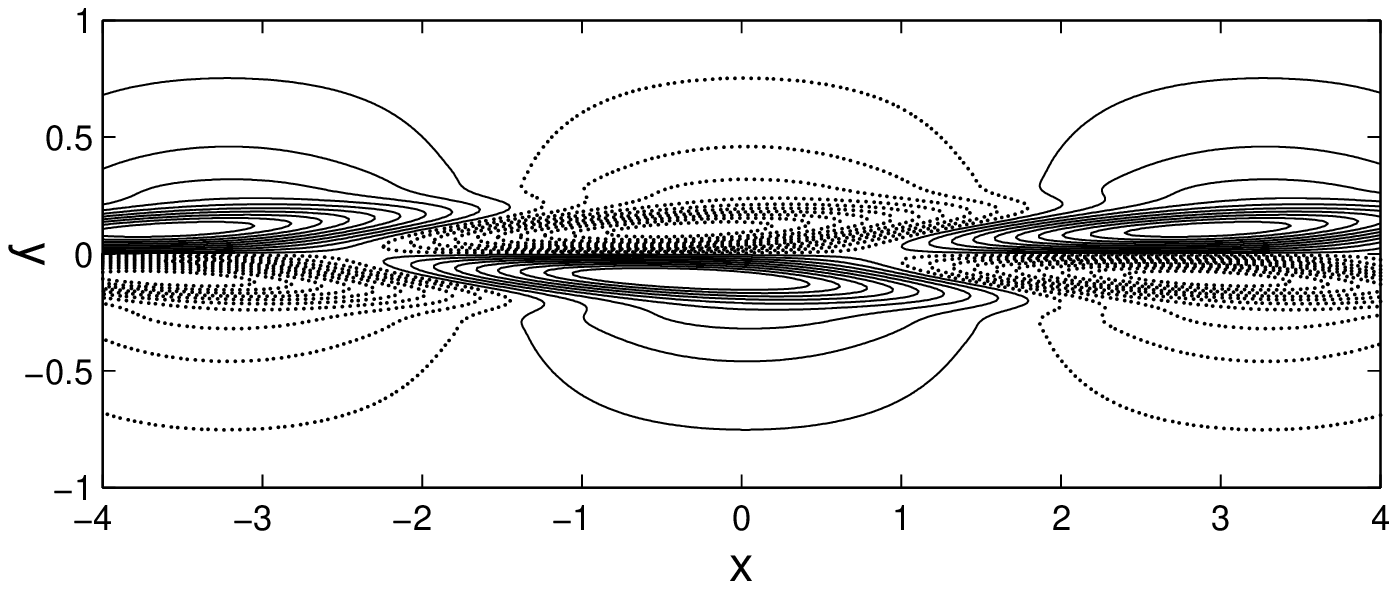}
\includegraphics[angle=90,width=3cm,height=12cm]{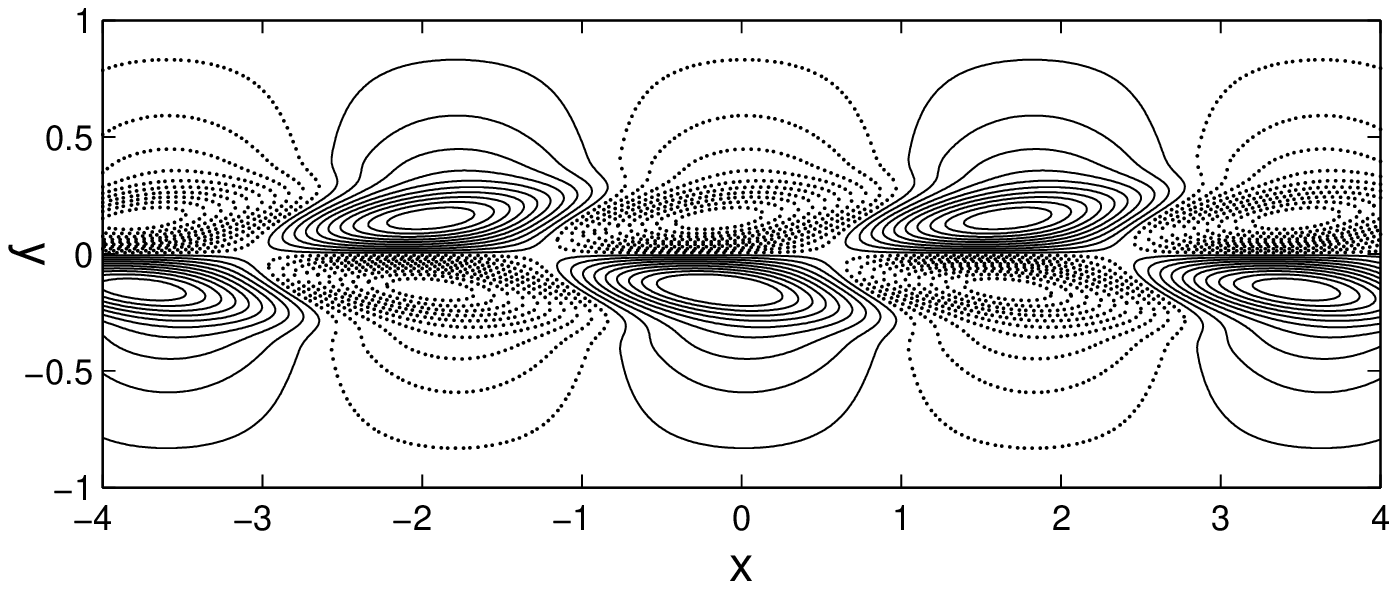}
\caption{Real part of the normalised streamfunction of the eigenmodes of 
the $-i\mathcal M^{-1} \mathcal L_{OS}$ operator for $H=10$ (left) and $H=100$ (right). Critical mode from the A branch (see figure \ref{fig:spectra}) (top) and least dissipative mode from the P branch (bottom). The dotted lines correspond 
to negative values of the streamfunction while solid lines represent positive values.}
\label{fig:eigmodes}
\end{figure}

\begin{figure}
\includegraphics[width=8.5cm]{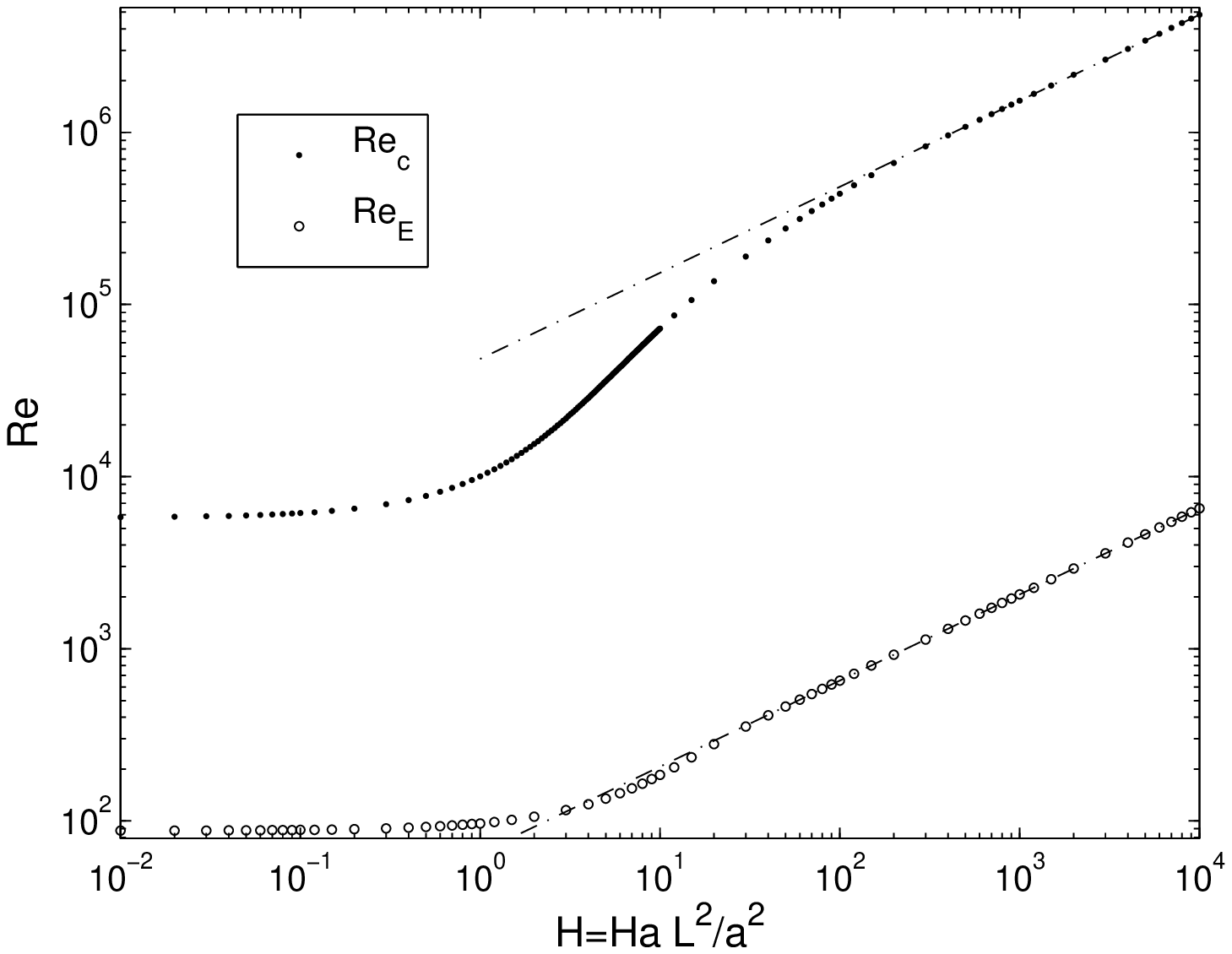}
\includegraphics[width=8.5cm]{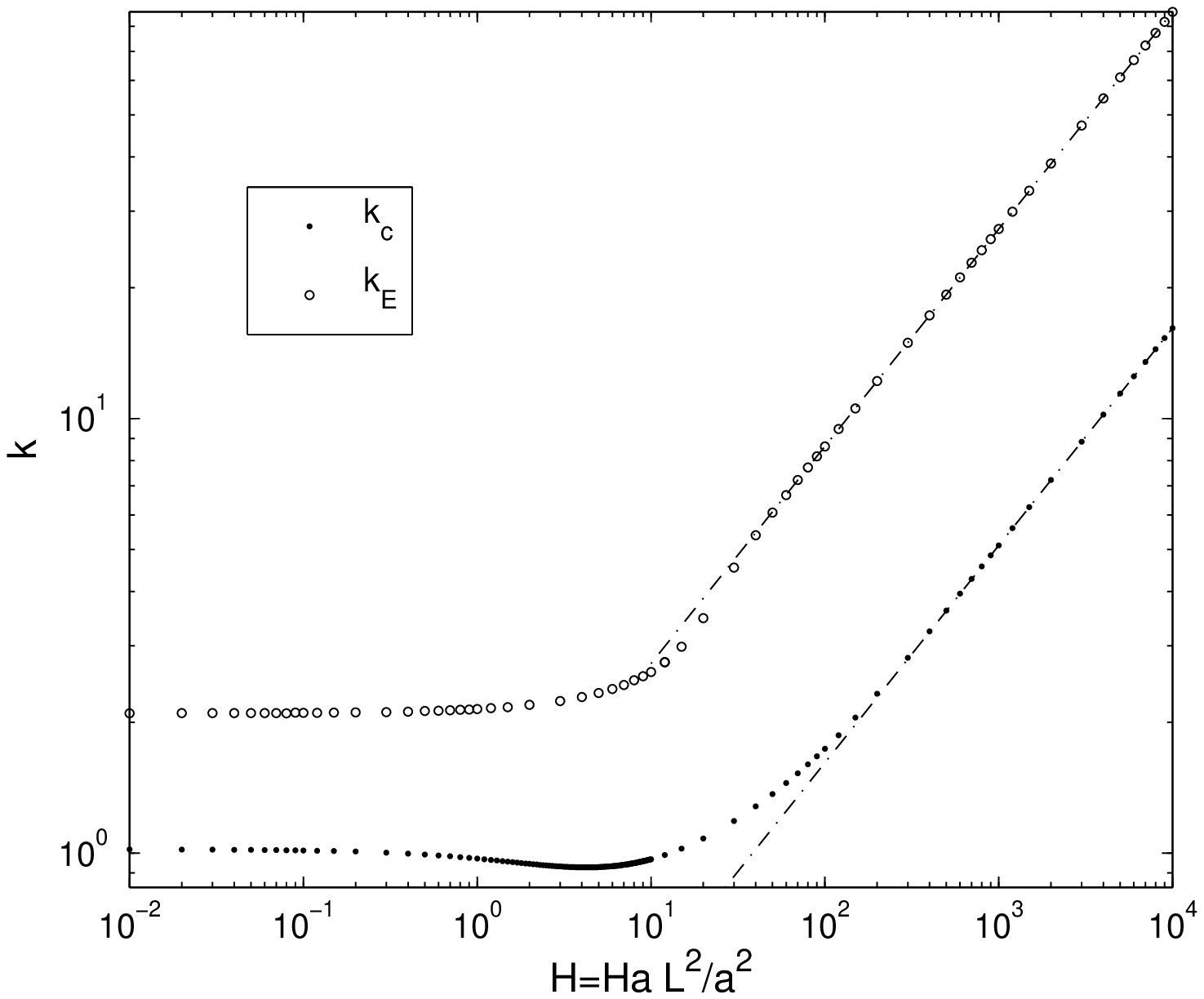}
\caption{Critical Reynolds number (top) and critical wavenumber (bottom) 
obtained from the linear stability analysis (section \ref{sec:linstab})  and 
the energy stability analysis (section \ref{sec:energy}). These two 
critical Reynolds 
numbers provide respectively an upper and a lower bound for the stability 
threshold. The points correspond to the effectively calculated values.}
\label{fig:rkc}
\end{figure}

Figure \ref{fig:spectra} shows the eigenvalue spectra of $\omega$ from the 
discretised 
operator $-i\mathcal M^{-1} \mathcal L_{OS}$ near criticality, for several 
values of $H$. The spectra exhibit
the same three branches as those labelled A, P and S by \cite{mack76} for the 
case of the plane Poiseuille flow ($H=0$). As $H$ 
 increases, the number of weakly dissipated modes along the P branch increases 
 faster  than that along the A Branch. These modes correspond to vortex patterns
developing in the centre of the channel (see figure \ref{fig:eigmodes}). 
However the first unstable mode to appear is always one of the A branch. This
critical mode is mostly the quasi-2D MHD equivalent of the Tollmien-Schlichting 
waves in 
Poiseuille flows. Additionally, a centre region of growing size and filling 
density when $H$ increases, appears at the junction of the A, P and S  branches. 
A similar behaviour has already been observed by \cite{airiau04} in the case of 
the 3D Hartmann channel flow problem. \cite{dongarra96} has shown that the 
appearance of this region at the branch junctions in the Orr-Sommerfeld problem 
associated to the Poiseuille flow was due to finite numerical precision. Since
(\ref{eq:os}) is formally identical to the Orr-Sommerfeld equation for the 
Poiseuille flow, as only the base profiles differ, this result also applies 
here. The affected eigenvalues however have a large negative imaginary part so 
they don't affect the stability results presented here. Therefore and in order 
to keep computational costs low, the numerical precision wasn't increased beyond
 64 bits.\\
The critical Reynolds and wavenumbers are plotted on figure \ref{fig:rkc}, 
together with their critical counterpart obtained from the energy stability 
analysis (see section \ref{sec:energy}). In the limit $H\rightarrow0$, we 
recover the critical values for the 2D Poiseuille flow, as mentioned in section 
\ref{sec:lsnum}. An asymptotic regime appears in the limit $H\rightarrow \infty$
(in practise for $H$ higher than around 200), for which 
$Re_c=4.83504 .10^4 H^{1/2}$ and $k_c=0.161532 H^{1/2}$. In this case, the 
boundary layers at $y=-1$ and $y=1$ don't interact with each other so the 
problem is governed by the stability property of each boundary layer. The 
lengthscale $L$ becomes irrelevant and the problem is governed solely by 
the Reynolds number based on the thickness of the Shercliff layer $Re/H^{1/2}$.
Interestingly, the critical wavenumber $k_c(H)$ reaches a minimum of 
$k^{min}_c=0.92736\pm8.10^{-5}$ for $H=4.2\pm0.1$, \text{i.e.} between these 
two asymptotic regimes.\\
High values of the critical Reynolds number scaled on the boundary layer
 thickness close to that found here also characterise the stability of 
 suction layers \cite{libby52} and Hartmann layers \cite{lock55,lingwood99}. 
 The reason for this similarity is that even 
 though the governing equations for these 3 problems differ (for example, the 
 governing equations for the Hartmann layer have an extra term involving the 
 electric potential not present in (\ref{eq:sm82})), all three boundary layers
 have an exponential base profile.

\begin{figure}
\includegraphics[width=8.5cm]{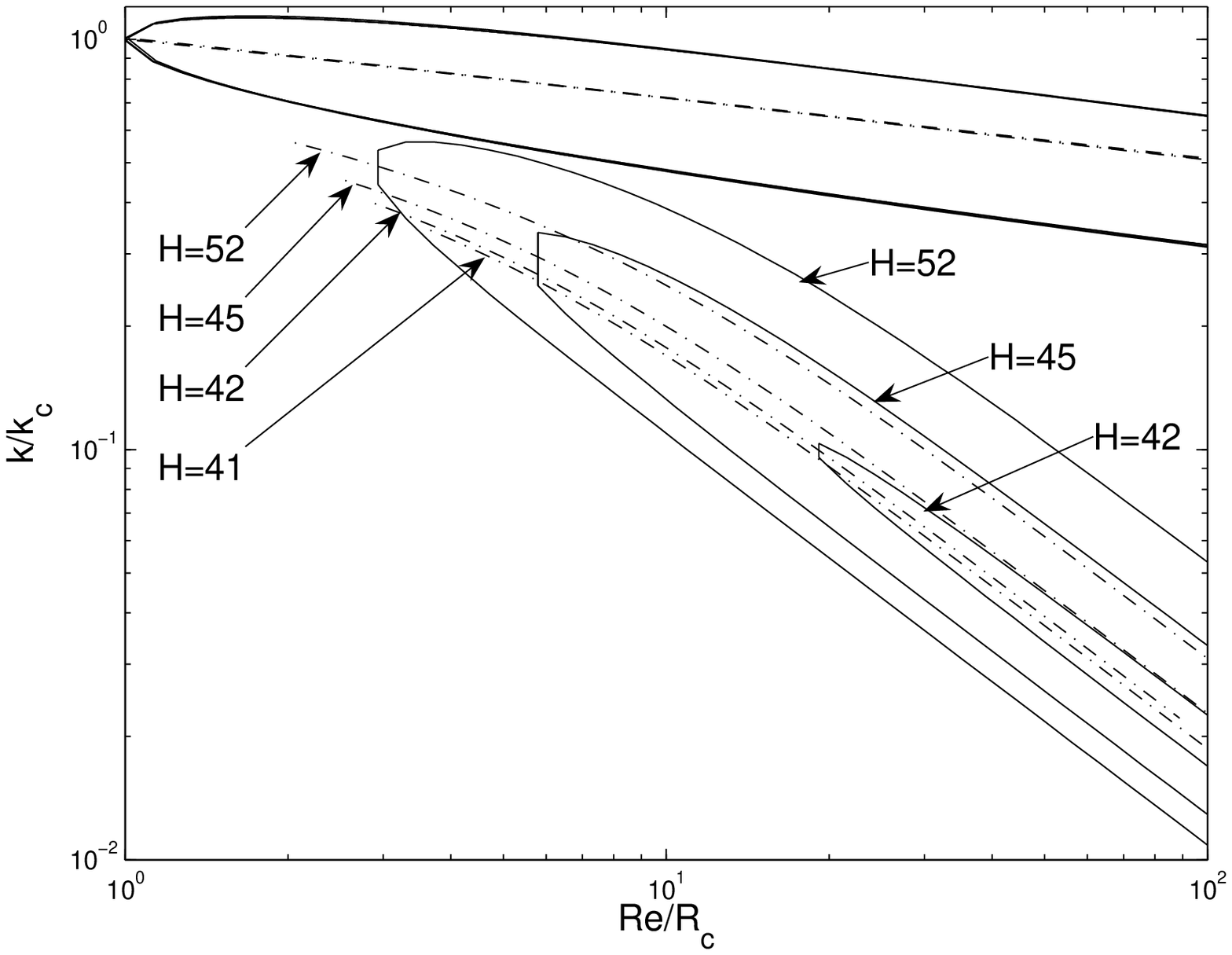}
\includegraphics[width=8.5cm]{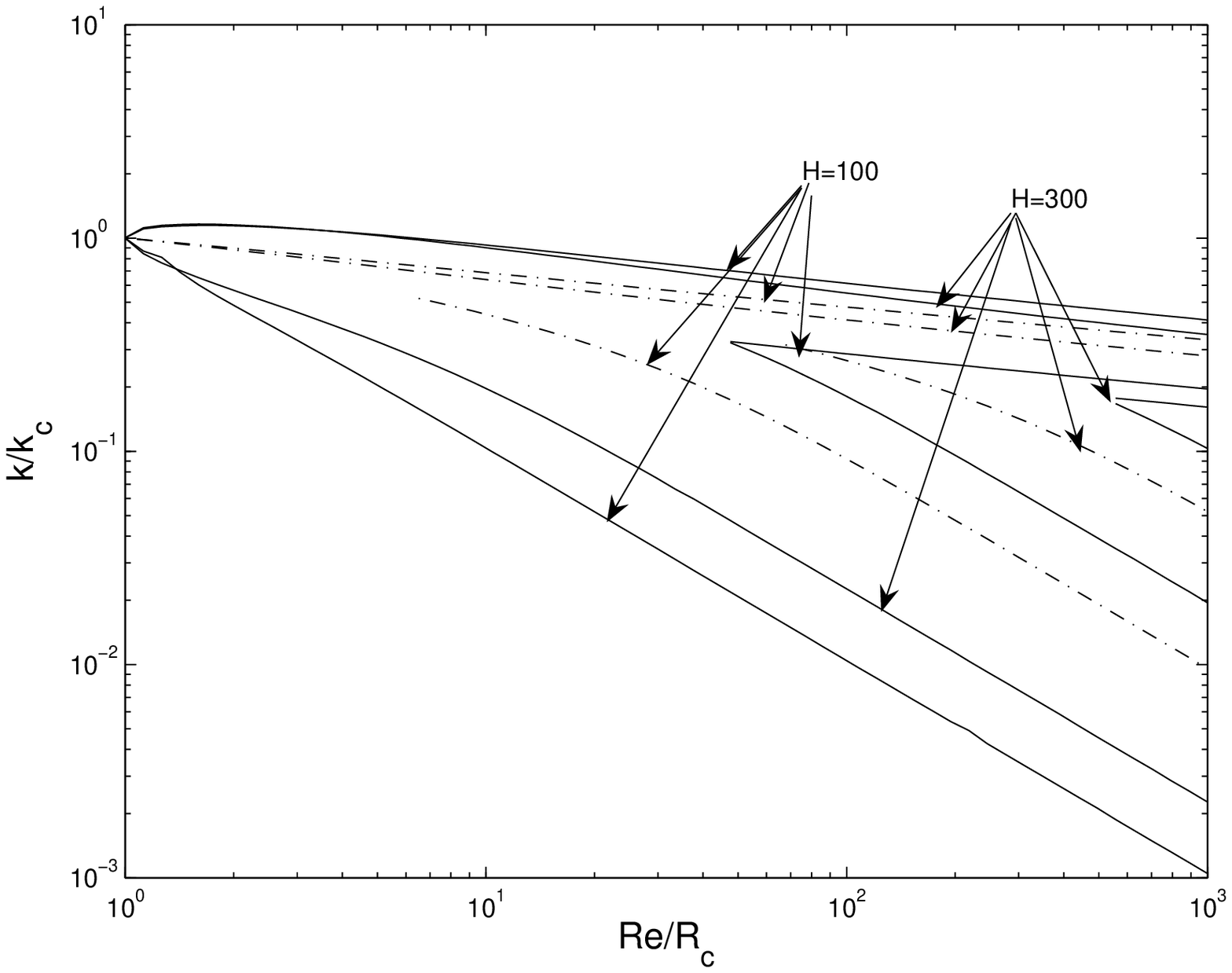}
\caption{Neutral stability curves (solid lines) and wavenumber of maximal
growthrate (dash-dot) as a 
function of the parameter $Re/Re_c$  several values of $H$. For low values of 
$H$ around 45, the two branches of instabilities are separated. The upper 
branch hardly varies between $H=41$ and $H=52$  whereas the lower one 
changes significantly (top).  The 
maximum of the 
growthrate due to the eigenvalue of the A2 branch exists also for values of 
$Re$ where it is not positive (also for $H<42$). For higher values of $H$, 
the two branches merge (bottom). For $H\sim 300$ the upper branch 
has reached  an asymptotic curve in the plane $(Re/Re_c,k/k_c)$, whereas the 
lower one hasn't.}
\label{fig:neutral_rk}
\end{figure}

\begin{figure}
\includegraphics[angle=90,width=3cm,height=12cm]{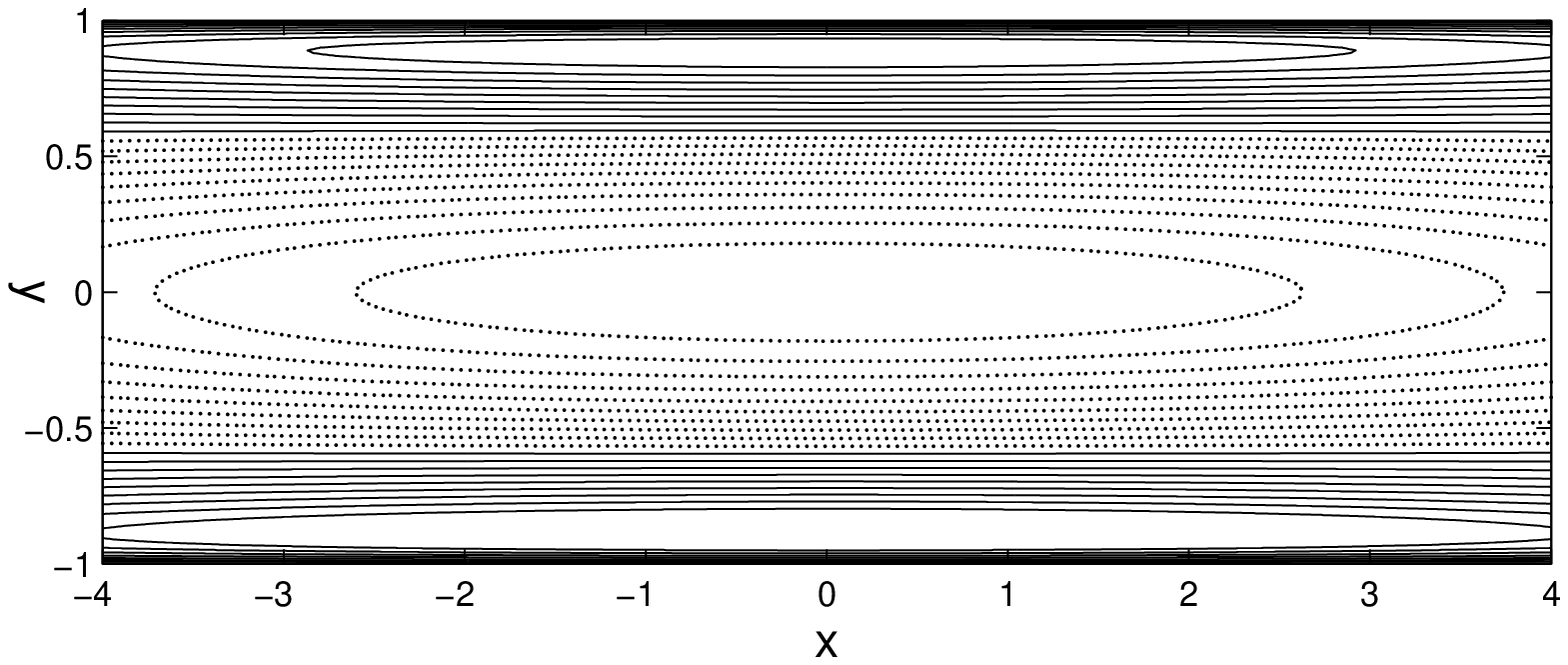}
\includegraphics[angle=90,width=3cm,height=12cm]{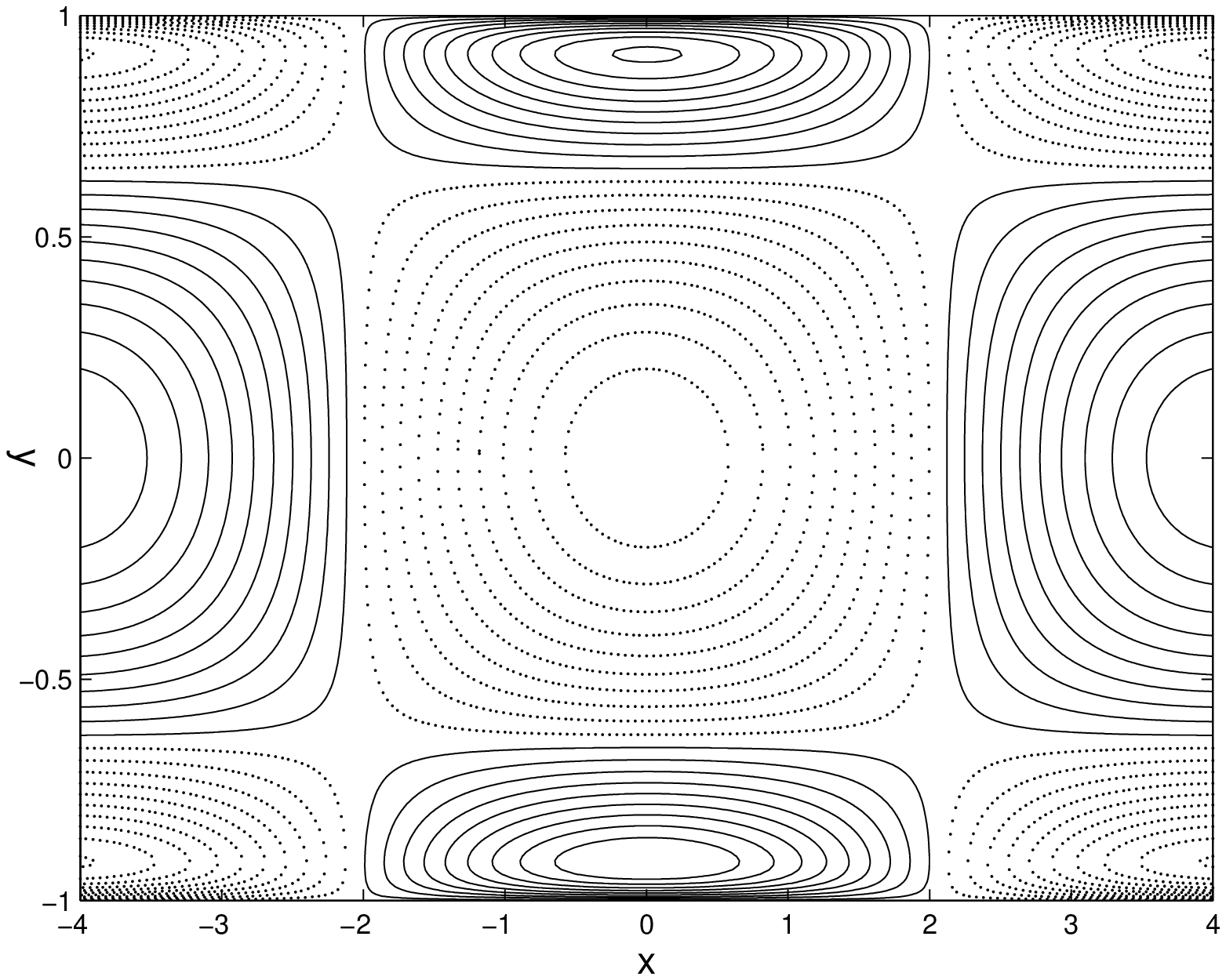}
\caption{Most unstable mode of the lower branch for $H=100$, $Re/Re_c=100$ 
(left, the corresponding wavelength is $k=0.09222k_c$, and the two branches are 
separated for this value of $Re$) and for $H=300$, $Re/Re_c=96.17$ (right, the 
corresponding wavelength is $k=0.02711k_c$, and the two branches are not 
separated for this value of $Re$, see figure \ref{fig:neutral_rk})}
\label{fig:eigmodes_lowb}
\end{figure}

A deeper insight into the linear stability of the flow can be gained with help
of the neutral curves which represent the border between stable and unstable 
modes in the $(Re/Re_c,k/k_c)$ plane. These are plotted on figure 
\ref{fig:neutral_rk}, 
along with the curves for the  wavelength achieving maximal growthrate. The 
first noticeable feature is that  the unstable
 modes are located in relatively narrow bandwidths which drift toward the
 lower wavenumbers as $Re$ increases. This rather contrasts with the large
  high pitched bandwidth over which transient growth occurs, as will be seen in
  section \ref{sec:transg}. The second noticeable feature is that the set of 
  unstable modes is split into 2 branches. The upper (\textit{resp.} lower) 
  branch stems from the most unstable mode of the A1 (\textit{resp.} A2) 
  subbranch of the eigenvalue spectrum (see figure \ref{fig:spectra}). The 
  modes from the upper branch are the Tollmien-Schlichting waves mentioned 
  earlier in this section the most unstable of which is always more unstable 
   than that of the lower one. The modes  corresponding to the lower branch 
   are made of one strong vortices in each of the boundary layers and one weaker 
   central one. In contrast to the antisymmetric critical modes of the upper 
   branch (see figure \ref{fig:eigmodes}), they are symmetric about the 
   centre of 
   the duct (figure \ref{fig:eigmodes_lowb}). At low values of $H$,  the lower 
   branch is separated 
   from the upper one and appears at increasing values of $Re$ and $k^{-1}$ 
   when $H$ decreases. We couldn't find this lower branch for $H<42$ but it is 
   difficult to 
   tell from our discrete numerical calculation whether it is there or not as 
   it could persist at 
   much higher values of $Re$ and $k^{-1}$.  When $H$ increases, the lower 
   branch merges into the 
   upper and keeps moving towards very high $Re$ whereas the upper branch 
   converges towards an asymptotic curve in the plane $(Re/Re_c,k/k_c)$. 
%
\section{Energy stability analysis}
\label{sec:energy}
\subsection{Principle and formulation}
We now seek a lower bound for the Reynolds number at which the 
base flow (\ref{eq:base}) becomes unstable. Such a lower bound is obtained
by looking for the maximum value of the Reynolds number $Re_E$ for which the 
energy 
$E=\|\tilde{\mathbf u}\|^2$ 
of any given perturbation $\tilde{\mathbf u}=\mathbf u-\mathbf U$ decays
 monotonically (here the 
functional norm is the usual $\mathcal L_2$ norm). Following \cite{joseph76},
we start from the equation governing the evolution of $E$, which is readily 
derived from (\ref{eq:sm82}):
\begin{eqnarray}
\frac{1}{2}\frac{dE}{dt} &=& -i\omega_E[\tilde{\mathbf u}]E\\
-i\omega_E[\tilde{\mathbf u}] &=& 
-\frac{\int \tilde{\mathbf u}\cdot\mathbf D[\mathbf U] \tilde{\mathbf u} d\Omega
+\frac{1}{Re}\|\nabla \tilde{\mathbf u}\|^2}{E}-\frac{H}{Re}
\end{eqnarray}
$\mathbf D$ denotes the deformation tensor based on the laminar solution 
(\ref{eq:base}): $D_{ij}=1/2(\partial_{x_i}U_{x_j}+\partial_{x_j}U_{x_i})$. 
Let now $S$ be the subspace of $\mathcal L_2$ spanned by solenoidal 
vector fields which satisfy the no slip boundary conditions at the side walls. 
Then $Re_E$ is the highest value of $Re$ such that the maximum of the 
functional 
$i\omega_E[\tilde{\mathbf u}]$ over $S$ is negative. This optimisation problem
is solved using variational calculus and introducing a Lagrange multiplier to 
enforce the constraint of mass conservation. After elimination of the latter,  
the maximum value of $\omega_E$ and the function which achieves it are found 
to be solutions of the following eigenvalue problem:
\begin{eqnarray}
\mathcal L_Ev &=& -2i\omega \mathcal M v
\label{eq:es}\\
\mathcal L_E&=& ik(U^{\prime\prime}+2\mathbf U^{\prime}D)
+\frac{2}{Re}\mathcal M^2-2\frac{H}{Re}\mathcal M
\nonumber
\end{eqnarray}
Contrarily to the eigenvectors of 
$i\mathcal M^{-1}\mathcal L_{OS}$, which are solution of the linearised motion
equations, and would therefore correspond to a possibly observable 
flow pattern if linear instability was the driving mechanism, 
those of $i/2\mathcal M^{-1}\mathcal L_E$ are no solutions of the motion 
equations. This partly explains that the lower bound given by the energy 
stability analysis often lays far below any observed instability threshold. It,
however, has the advantage of not relying on any assumption made either on the 
equations or on the perturbations. In particular, finite amplitude 
perturbations are counted in. Here again, the introduction of the frictionless
 growthrate $\omega_{E0}=\omega_E-\frac{H}{iRe}$ makes (\ref{eq:es}) formally 
 identical to the hydrodynamic Poiseuille problem.\\

\subsection{Results}
\begin{figure}
\begin{center}
\includegraphics[angle=90,width=3cm,height=12cm]{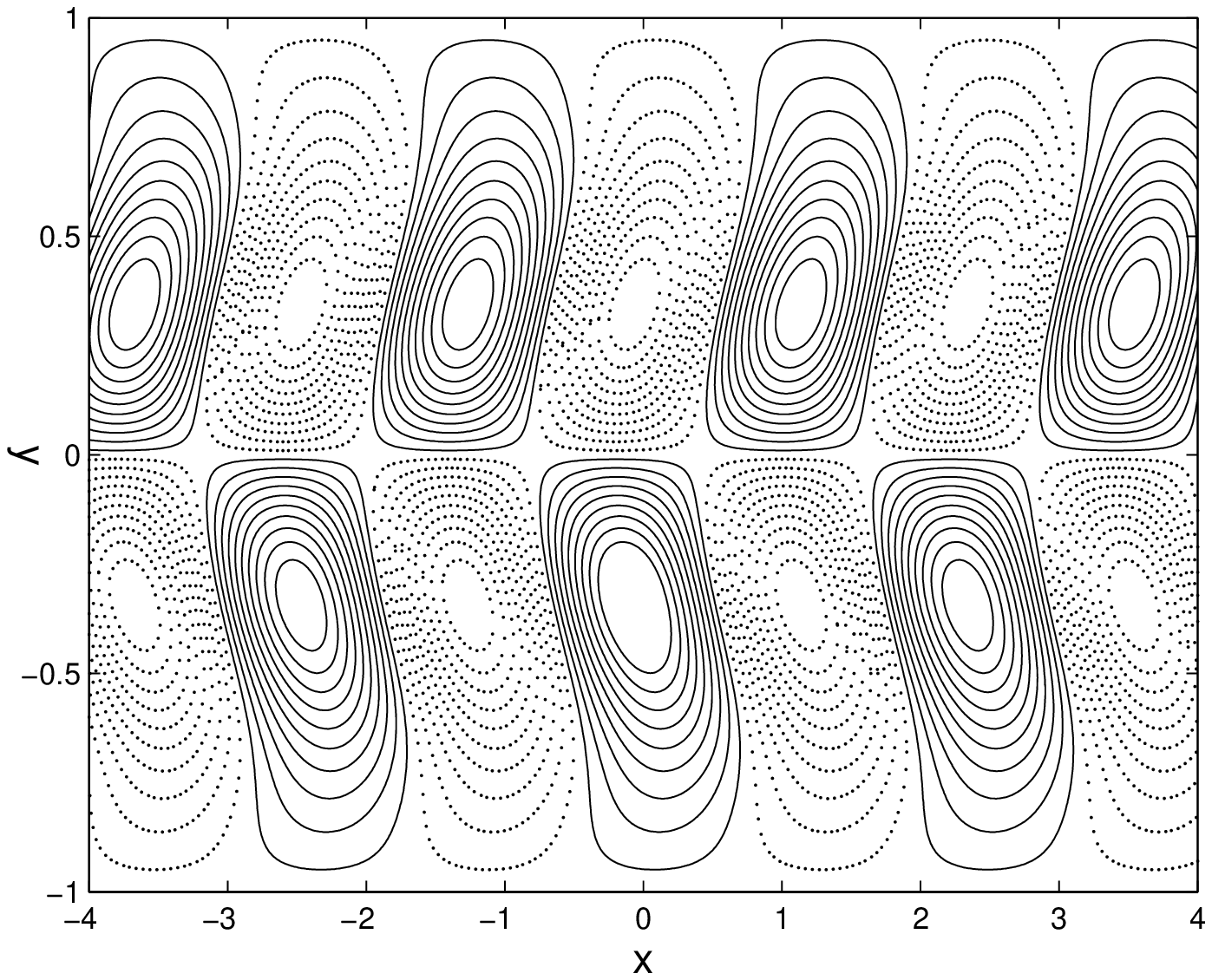}
\includegraphics[angle=90,width=3cm,height=12cm]{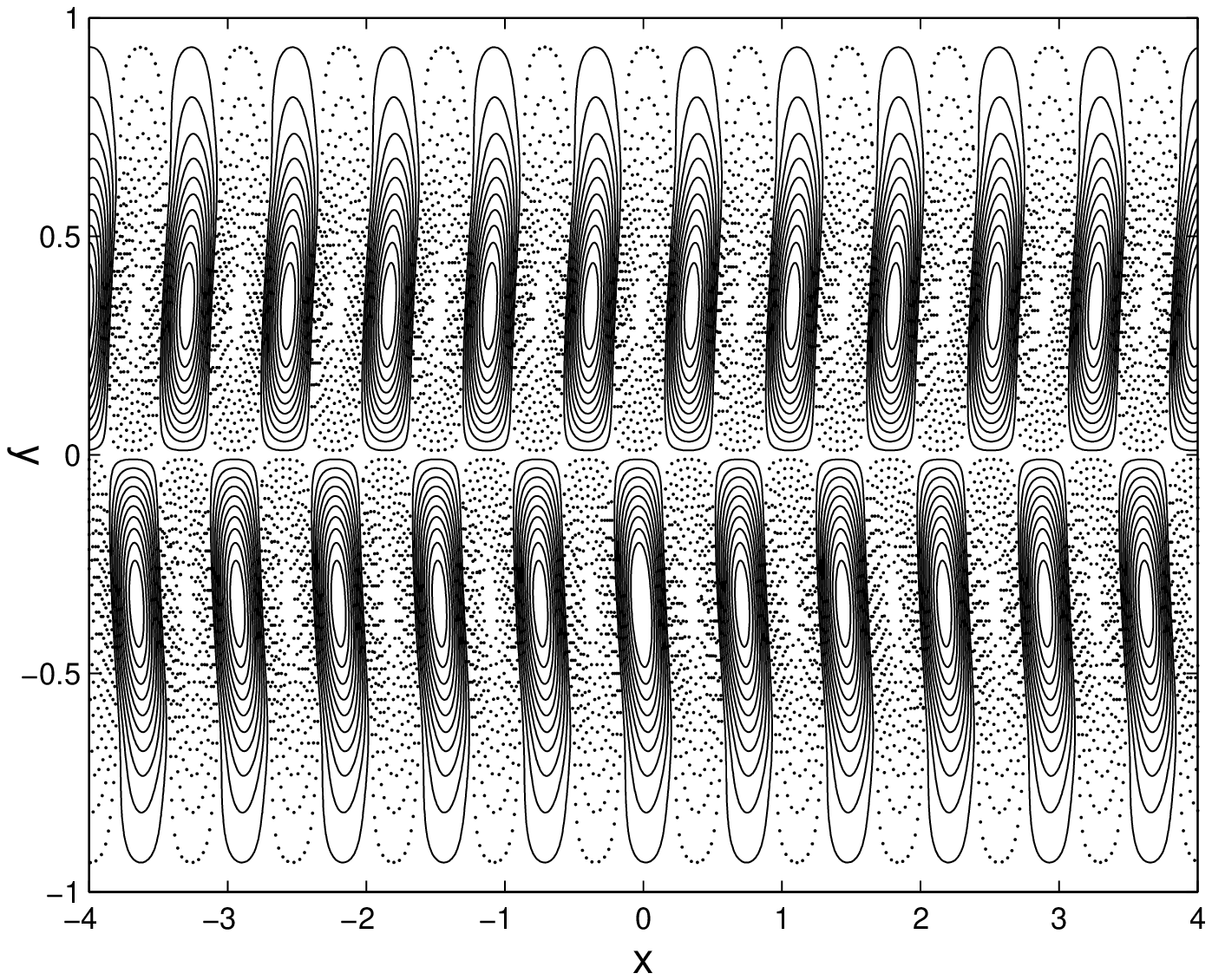}
\caption{Critical modes for the energy stability for $H=10$ (left, $Re_E=1.84703 .10^2$,
$k_E=2.61287$), $H=100$ (right, $Re_E=6.53225.10^2$, $k_E=8.62501$).}
\label{fig:critmod_e}
\end{center}
\end{figure}

Since the eigenvalue problems (\ref{eq:os}) and (\ref{eq:es})  only differ 
through the expression of the linear 
operators involved, (\ref{eq:es}) is solved using the same numerical
method and procedure as for the linear stability analysis (see section 
\ref{sec:lsnum}).\\
A distinctive feature of the 2D energy stability analysis, as opposed to the 
3D one is that there cannot be any solenoidal perturbation satisfying $k=0$. 
In the 3D channel flow problem without magnetic field, \cite{busse72} and 
\cite{joseph76}  have shown 
that these streamwise-independent perturbations are precisely those 
achieving the maximum energy growthrate. This remarkable property can therefore 
not be extended to flows governed by a 2D equation.\\
%
%
The spectra of eigenvalues of $i/2\mathcal M^{-1}\mathcal L_E$ possess only one branch and 
the related eigenfunctions correspond to vortex-like patterns regularly 
spread across the channel width. The number of vortices increases as the 
eigenvalue tends to $-\infty$. The critical mode is always made of two 
rows of anti-symmetric vortices as shown on figure \ref{fig:critmod_e}.
The dependence on $H$ of the critical Reynolds $Re_E$ and streamwise wavenumber 
$k_E$ for the energy stability analysis are shown on figure \ref{fig:rkc}. 
Contrarily to the critical wavelength for the linear stability, that for 
the energy stability increases monotonically with $H$. In the limit 
$H\rightarrow\infty$, the energy stability threshold for the flow is that 
of a single side boundary layer, characterised by the single critical parameter 
$Re_E/H^{1/2}=65.3288$. The related wavelength of the critical mode tends to 
$k_E=0.863484 H^{1/2}$.
\section{Transient growth}
\label{sec:transg}
\subsection{Principle and formulation}
In the two previous sections, we have found an upper and a lower bound 
for the threshold of the instability occurring in the channel. The lower 
bound has been obtained as a "no perturbation growth" condition, as all 
perturbations decay monotonically for $Re<Re_E$. The fact that it departs
 strongly from the upper bound, which corresponds to a condition for 
 infinitesimal perturbations to grow, leaves open the possibility for some 
 perturbations to
  still grow between these two extremes. Such a growth is known to stem from the 
  non-normality of the Orr-Sommerfeld operator. For such operators, a 
  linear combination of eigenmodes of the operator can undergo a short but 
  possibly intense transient growth, even though each normal mode monotonically
  decays. \cite{krasnov04} have shown that in the Hartmann layer problem, 
  such grown perturbations can act as  
  finite amplitude disturbances and destabilise the mean flow well below the 
  linear stability threshold. Since this mechanism is likely to also play a 
  key role in our problem, we shall now complete our overview of the 
  quasi-2D stability of the MHD duct flow by estimating the maximum transient 
  growth associated to the two-dimensional dynamics of equation (\ref{eq:sm82}).\\
  The detailed method to find the maximal transient growth and the associated 
  perturbations can be found for instance in \cite{schmid01}. It consists of 
  solving the non-modal linearised perturbation equation:
\begin{eqnarray}
\mathcal L_{OS}v &=&\mathcal M \partial_t v
\label{eq:os_nonmodal}
\end{eqnarray}
by expanding the perturbation over the set of Orr-Sommerfeld modes, with 
a vector of associated time-dependent coefficients 
$\mathbf \kappa=(\kappa_1,..\kappa_N)$, with $N$ 
large enough. The so-discretised solution $v_N$ of (\ref{eq:os_nonmodal}) then 
expresses as a function of a the diagonal matrix built from the eigenvalues of 
the Orr-Sommerfeld operator $\mathbf \Lambda=\diag(\omega_1,..,\omega_N)$:
\begin{equation}
\mathbf \kappa(t)=\mathbf \kappa(t=0)\exp(-it \mathbf \Lambda)
\end{equation}
Defining the maximum gain reached at time $t$ over the set of possible initial 
perturbations as 
$G(t)=\max\frac{\|\kappa(t)\|}{\|\kappa(t=0)\|}$, then $G(t)$  expresses as the 
square of the principal singular value $\sigma_1(t)$ of the matrix 
$C=F\exp(-it\Lambda)F^{-1}$, 
where $\mathcal M=F^HF$ is the Hermitian decomposition of $\mathcal M$. 
The left singular vector associated to $\sigma_1(t)$ 
represents the normalised  initial perturbation achieving maximum growth
at time $t$ (thereafter called optimal perturbation at time $t=0$), and the 
right singular vector associated to $\sigma_1(t)$
represents the vector field into which this same perturbation evolves 
at time $t$ (thereafter optimal perturbation at time $t$). It should also be 
underlined that $G(t)$ doesn't represent the time-evolution of any one 
perturbation but rather the envelope of the family of curves representing the 
evolution of all perturbations becoming at some time the most amplified.\\
\subsection{Numerical procedure}
We solve the problem numerically using the same expansion in Tchebychev 
polynomials as for the linear and energy stability analysis (see section 
\ref{sec:lsnum}). The singular value decomposition is performed with the 
standard MATLAB routine. This yields for given values of $H,Re$ and $k$, the 
growth $G(t)$ of the optimal perturbation at time $t$.\\
These calculations are repeated for $H\in\{0,1,3,10,30,100,300,1000\}$. For
each value of $H$, we take 10 values of $Re$ in geometric progression between 
$Re_E(H)$ and 
$Re_c(H)$ and 10 in geometric progression between $Re_c(H)$ and $100Re_c(H)$ in 
order to reach clean asymptotics. $k$ is taken in the interval $[0, 6k_c]$.  
Since we are interested in the maximum transient growth over time, we start by 
calculating 10 values of $G(t)$ and then refine the calculation around the 
maximum until a relative precision of $10^{-3}$ over $G_{mk}=\max_t G(t)$, and
$10^{-3}$ over $t_G$  (defined by $G(t_G)=G_{mk}$) is reached. The $G_{mk}$ 
maxima 
are gathered to provide a $G_{mk}(k)$ profile for each value of $(H,Re)$ and 
the same procedure as above is used to determine the maximum growth
 over $k$, $G_{max}$, and associated wavenumbers $k_{Gmax}$ and time $t_{Gmax}$.\\
Additionally, the number of modes used for the linear and energy stability (see section \ref{sec:lsnum})  turned out
to be insufficient to resolve the strong shear of the optimal modes 
in supercritical regime. The resolution had to be multiplied by $1.2$ for  
$Re>Re_c$ and progressively increased up to a factor $2.5$ for $Re=100Re_c$ 
in order to keep 
the variations of $G_{mk}$ and $G_{max}$ with the number of modes under the 
prescribed precision.
\subsection{Optimal modes given $H$, $R$, $k$ and $t$}
\begin{figure}
\includegraphics[width=8.5cm]{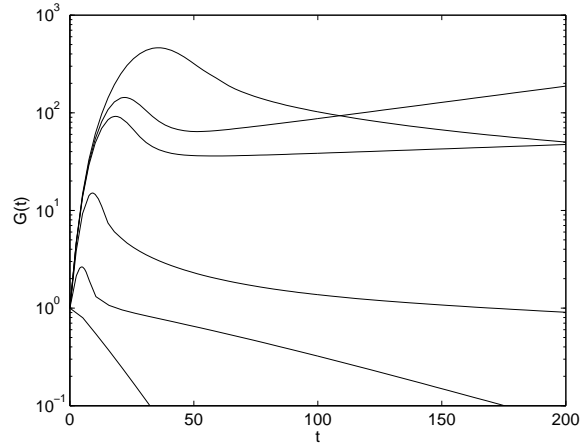}
\caption{$G(t)$, for $H=100$, $Re<Re,Re/Re_c=0.01, 0.1, 1.1, 2, 10$ 
(by order of 
growing maximum). For $Re<Re_c$, no growth occurs. For $Re/Re_c=1.1$ and $2$, the 
Tollmien-Schlichting 
wave of streamwise wavelength $k_c$ grows exponentially. This however happens  
quite a long time after a combination of modes of streamwise wavenumber $k_c$  
undergo
some significant transient growth.  For $Re/Re_c=10$, the normal modes of 
wavelength $k_c$ are no longer in the unstable bandwidth, but a combination of 
them still undergoes transient growth.}
\label{fig:gt}
\end{figure}

\begin{figure}
\includegraphics[angle=90,width=3cm,height=12cm]{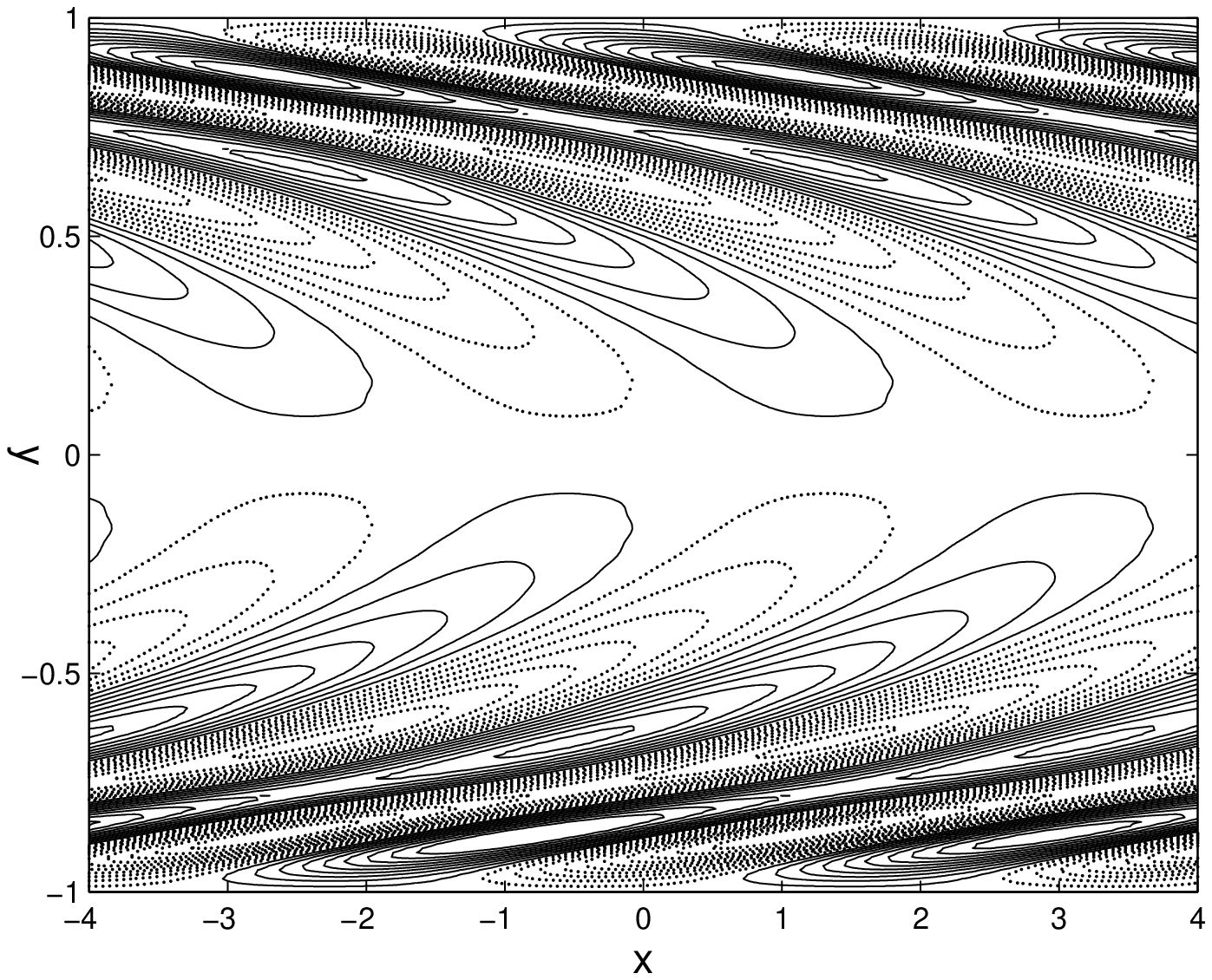}
\includegraphics[angle=90,width=3cm,height=12cm]{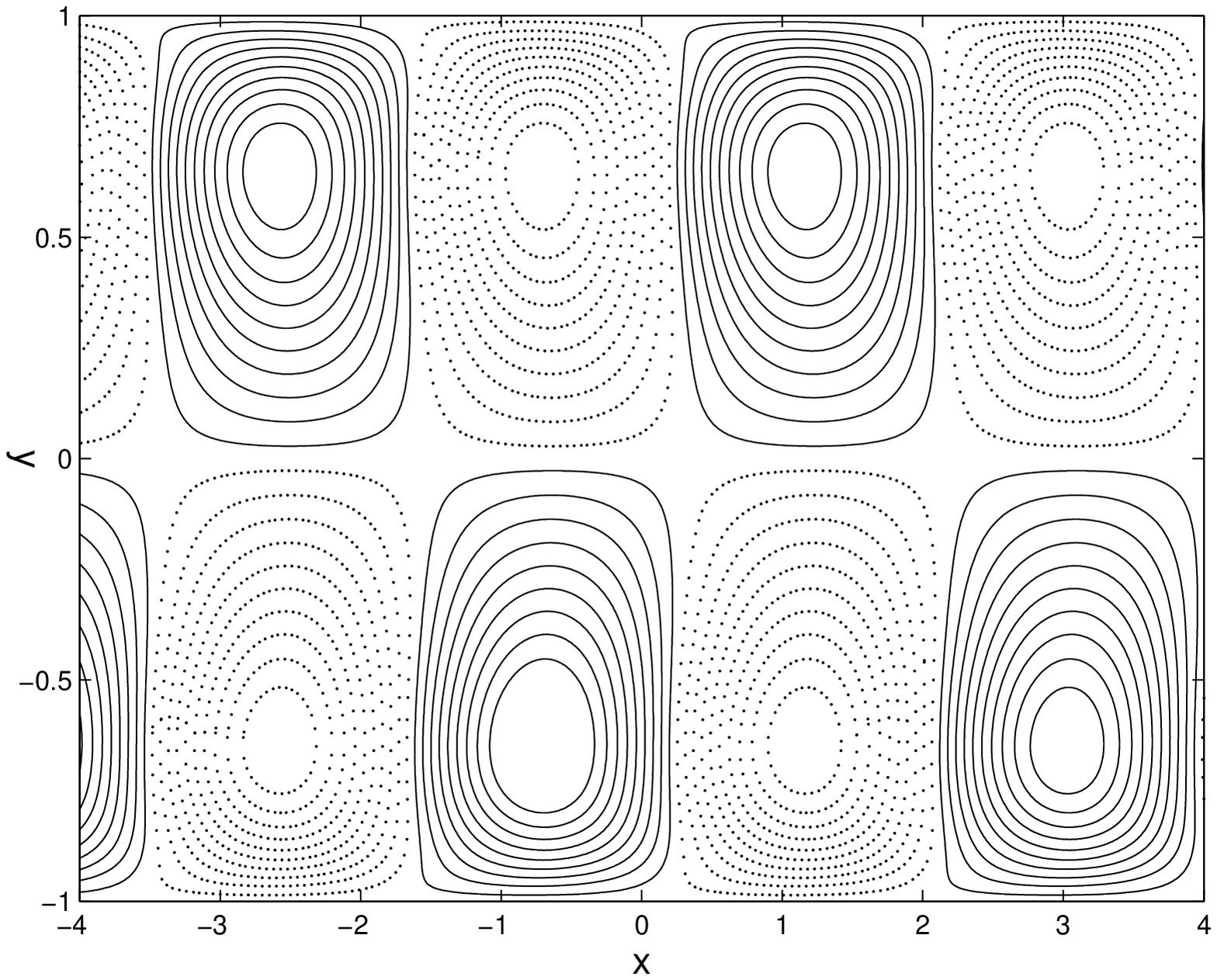}\\
\includegraphics[angle=90,width=3cm,height=12cm]{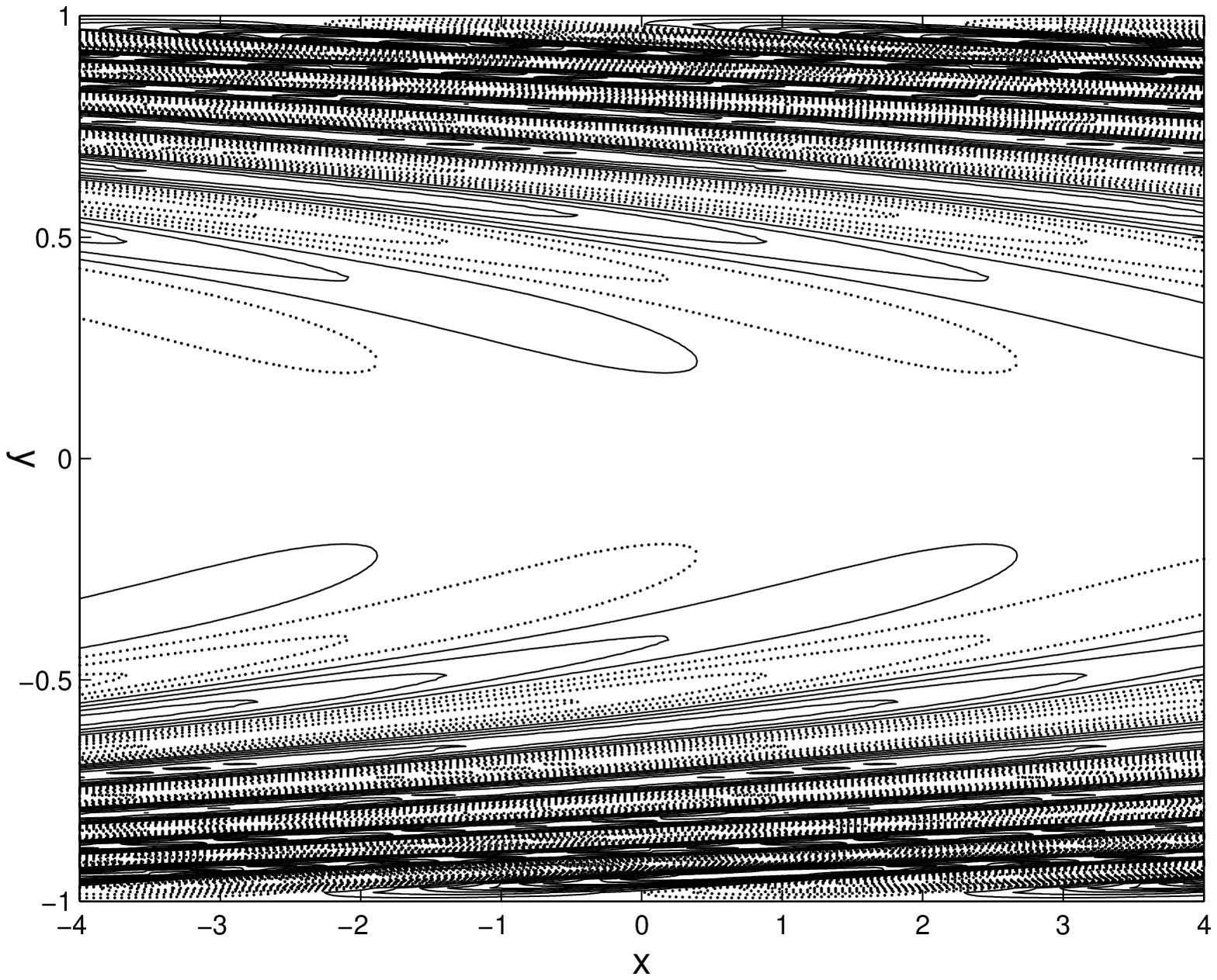}
\includegraphics[angle=90,width=3cm,height=12cm]{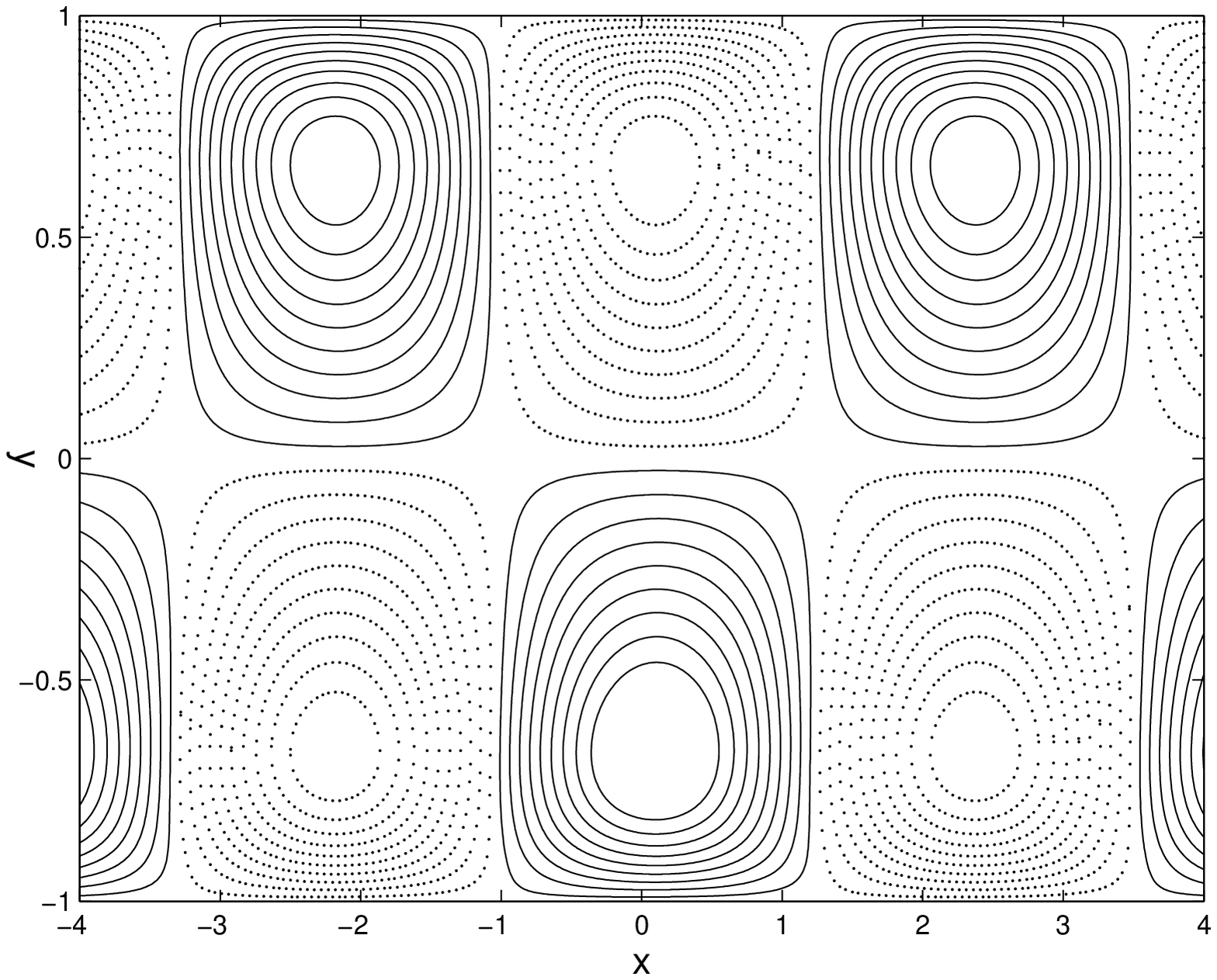}
\caption{Streamlines of the optimal perturbation achieving $G_{max}$ for $H=10$, $Re/Re_c=0.1366$ (top), $Re/Re_c=2.783$ (bottom). In both cases, the mode with 
strong  shear (left) represents the perturbation at $t=0$ and the flow on the 
right represent the same perturbation at $t=t_{Gmax}$(right).}
\label{fig:optmod10}
\end{figure}

\begin{figure}
\includegraphics[angle=90,width=3cm,height=12cm]{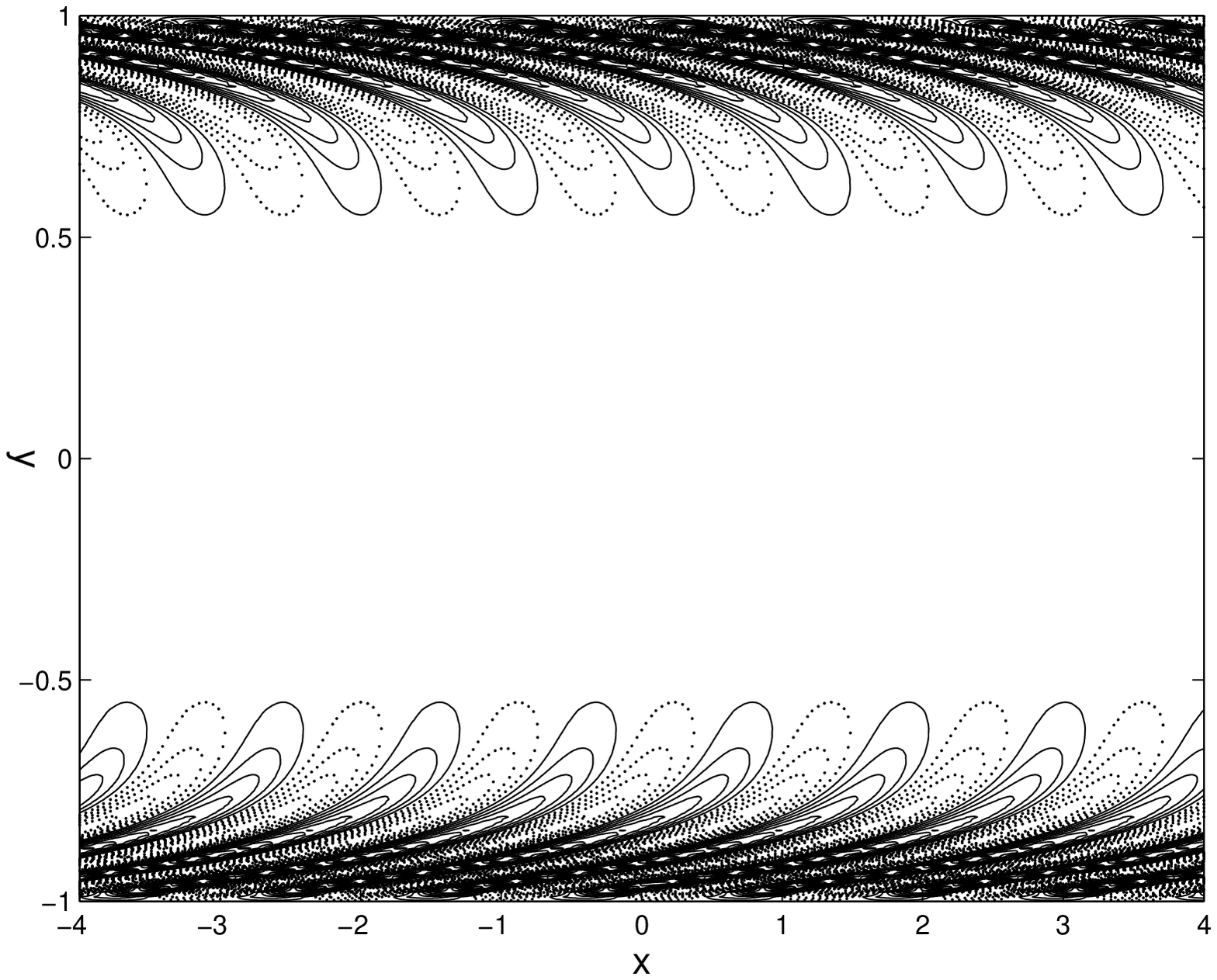}
\includegraphics[angle=90,width=3cm,height=12cm]{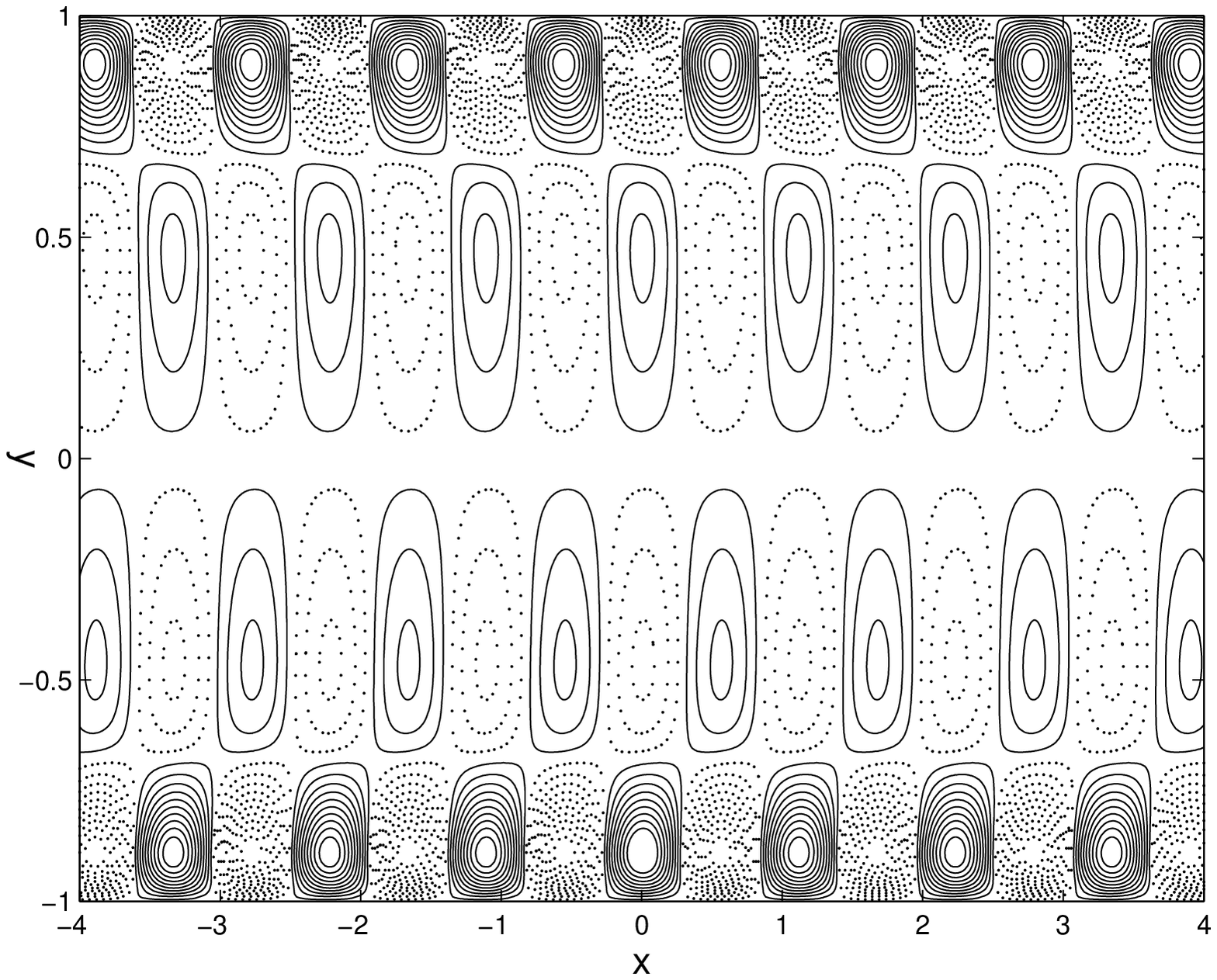}\\
\includegraphics[angle=90,width=3cm,height=12cm]{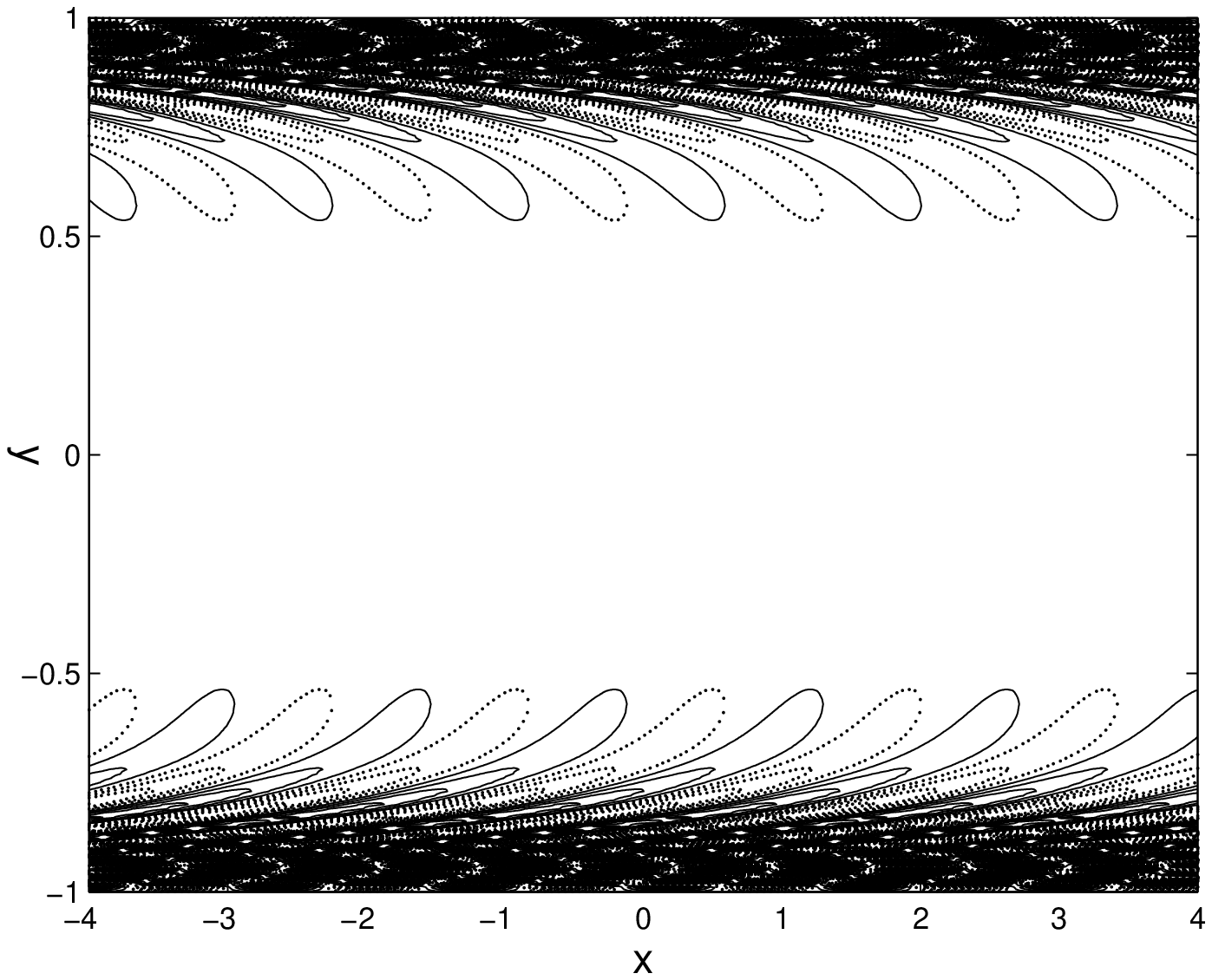}
\includegraphics[angle=90,width=3cm,height=12cm]{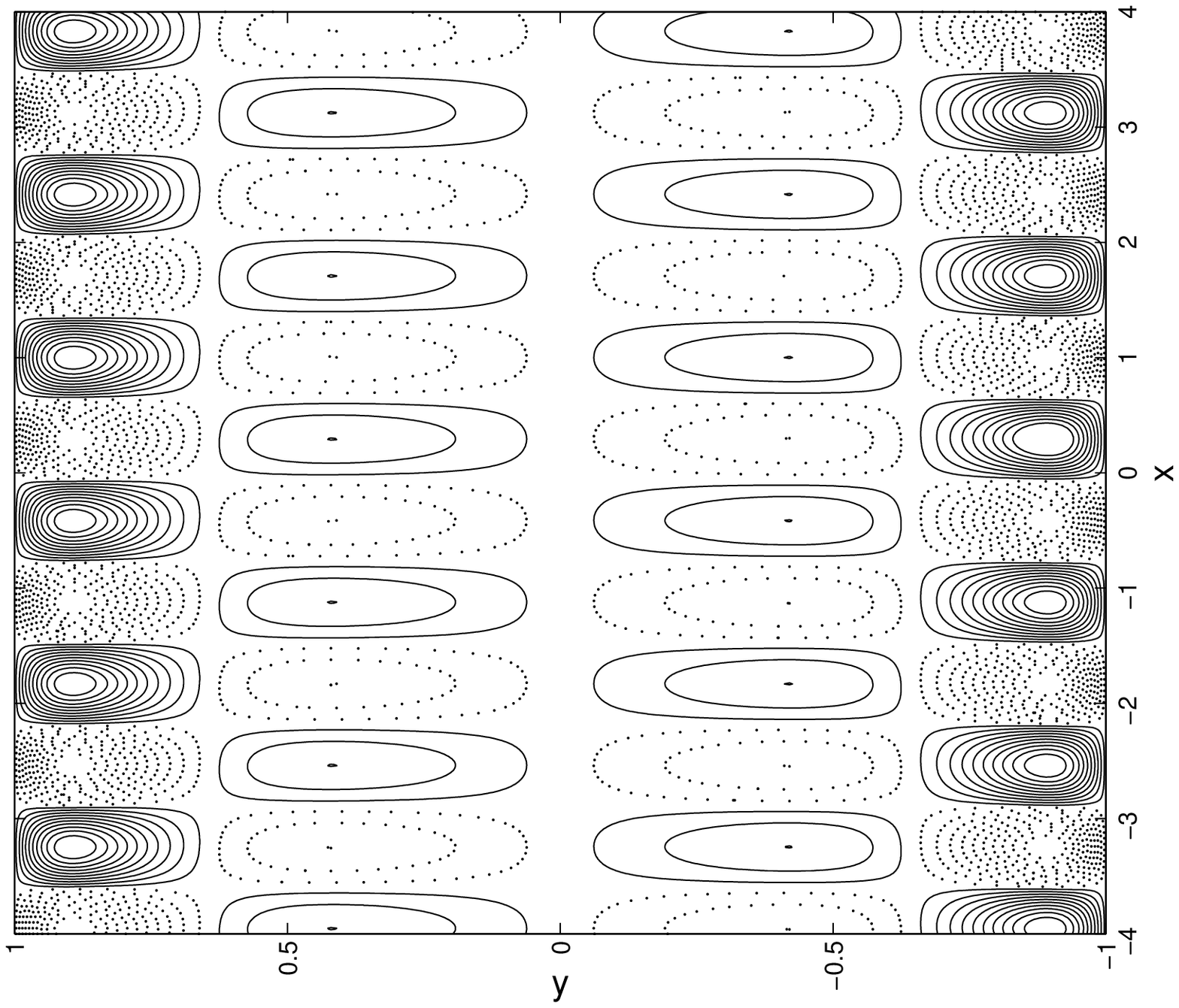}
\caption{Streamlines of the optimal perturbation, $H=100$, $Re/Re_c=0.1141$ (top) 
and $Re/Re_c=2.783$ (bottom). Optimal perturbation at $t=0$ (left) and 
$t=t_{Gmax}$(right).}
\label{fig:optmod100}
\end{figure}

The evolution of $G(t)$ is found to be qualitatively similar to that known for 
the two-dimensional Poiseuille flow (see for instance \cite{schmid01} p115). 
For $Re<Re_E$, since the energy stability returns a no growth condition, G(t) 
monotonically
 decays from $G(t=0)=1$ to $0$ when $t\rightarrow\infty$, and this for all 
values of $k$. For $Re_E<R<Re_c$, there exists some values of $k$ 
for which $G(t)$  first increases, reaches a maximum at $t=t_G$ and then decays
to $0$. For $Re>Re_c$, 
$G(t)$ still exhibits a local maximum but diverges when $t\rightarrow\infty$ 
for the wavelengths with linearly unstable normal modes. This means that 
there are modes undergoing transient 
growth as soon as $Re>Re_E$ and that even for $Re>Re_c$ transient growth exists 
and happens at a much earlier time than any substantial exponential growth from
the linearly unstable modes identified in section \ref{sec:linstab}. Some 
examples of these different cases are gathered on figure \ref{fig:gt}.\\
The optimal modes are plotted on figure 
\ref{fig:optmod10} and \ref{fig:optmod100}. As in the case of the energetic 
and linear stability, the 
scale of the transversal velocity variations follows that of the Shercliff
 layer 
thickness, so that both at $t=0$ and at the time of maximal growth $t_G$, the 
optimal perturbation presents an increasing number of vortices along the 
spanwise direction as $H$ increases. When the Reynolds number is increased, 
the streamlines of the optimal perturbation at $t=0$ undergo a stronger shear 
from the base flow so that for high enough values of $Re$, vortices degenerate
 in streaks of alternate vorticity which become more and more aligned with the 
streamwise direction as $Re$ is further increased.\\
Remarkably, at $t=t_G$, the optimal perturbation always evolves into modes which 
strongly resemble the critical normal modes found in section 
\ref{sec:linstab}. The 
general aspect of these optimal perturbations is very similar to that of those 
found  in the case of the infinite channel with spanwise 
magnetic field parallel to  the walls: by searching the optimal perturbations with possible
dependence in the 3 spatial directions it is found that for $Ha\gtrsim 100$, the 
optimal perturbations are 
2D and aligned with the magnetic field. For lower values of $Ha$, 
although these perturbations do undergo some transient growth, other 
perturbations with wavevector of non-zero a streamwise component  
(therefore three-dimensional) 
achieve the maximum growth. The strong similarity between
 the two problems suggests that in the limit of large $H$, the quasi-2D 
optimal perturbations found in this section for the duct problem 
 may well undergo a stronger transient growth than any other possible 3D 
 perturbations not
 taken into account in the present work. Of course, this remains to be proved -
 or disproved -  by a 3D analysis in the exact duct configuration.
\subsection{Maximum transient growth over $k$ and over time, for given $H$ and 
$Re$}
\begin{figure}
\includegraphics[width=8.5cm]{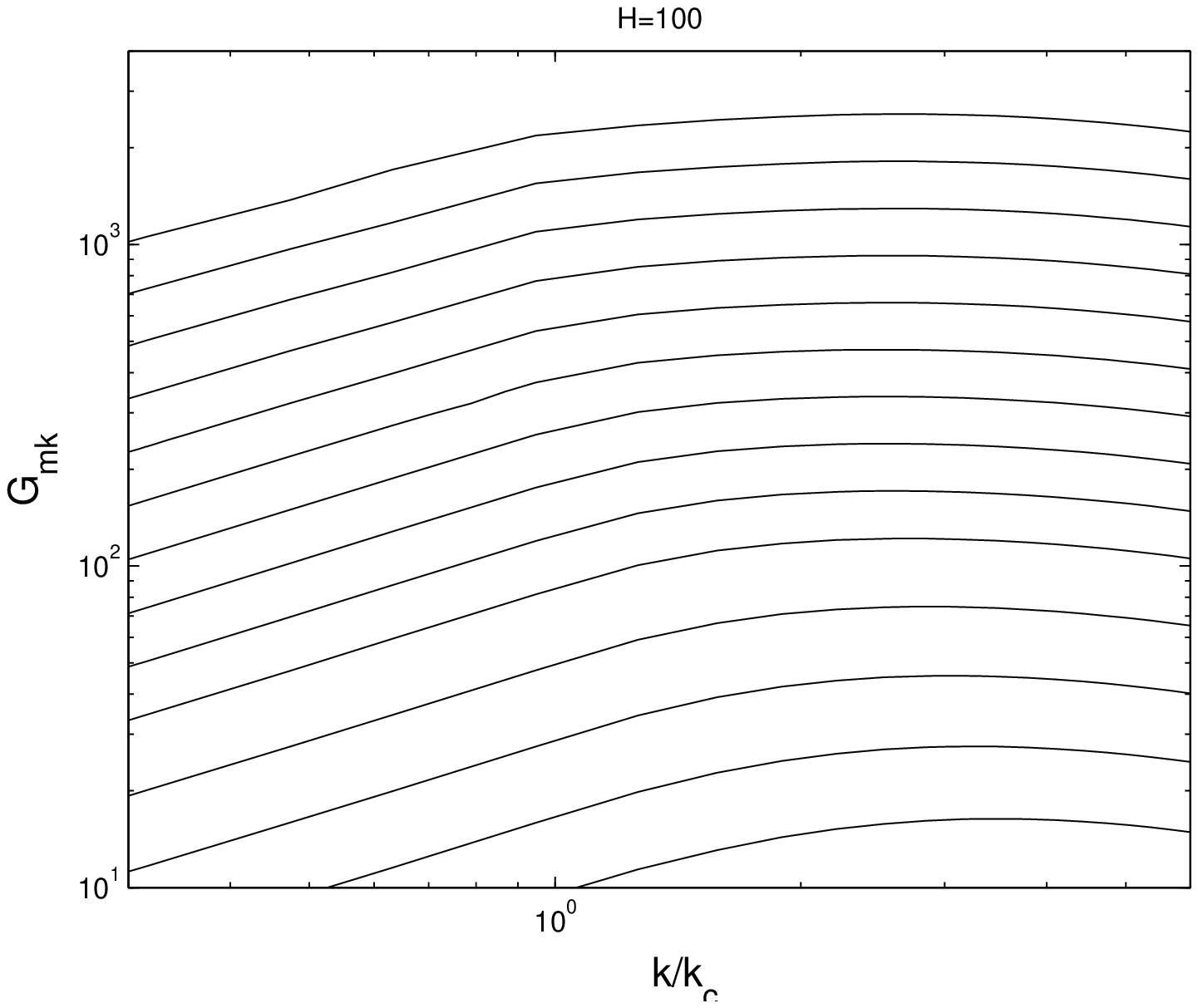}
\includegraphics[width=8.5cm]{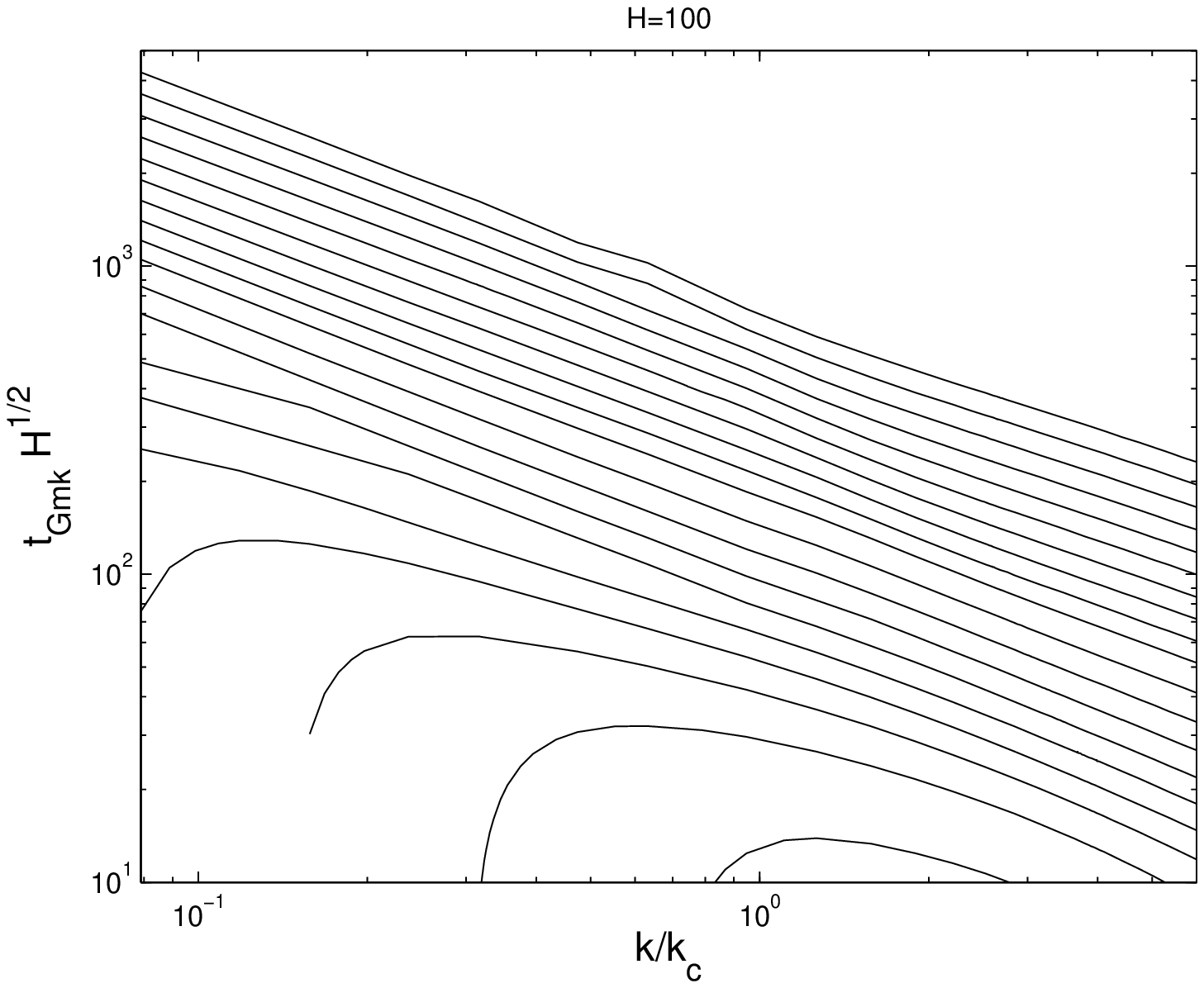}
\caption{Maximum transient growth (top) and time of maximum transient growth 
(bottom)  as a function of the streamwise wavenumber $k$, for $H=100$ , and for 
10 values of $Re$ in 
geometric progression between  $Re_E(H=100)$ and $Re_c(H=100)$ and 10 between 
$Re_c(H=100)$ and $100Re_c(H=100)$.}
\label{fig:gmax_k}
\end{figure}

\begin{figure}
\includegraphics[width=8.5cm]{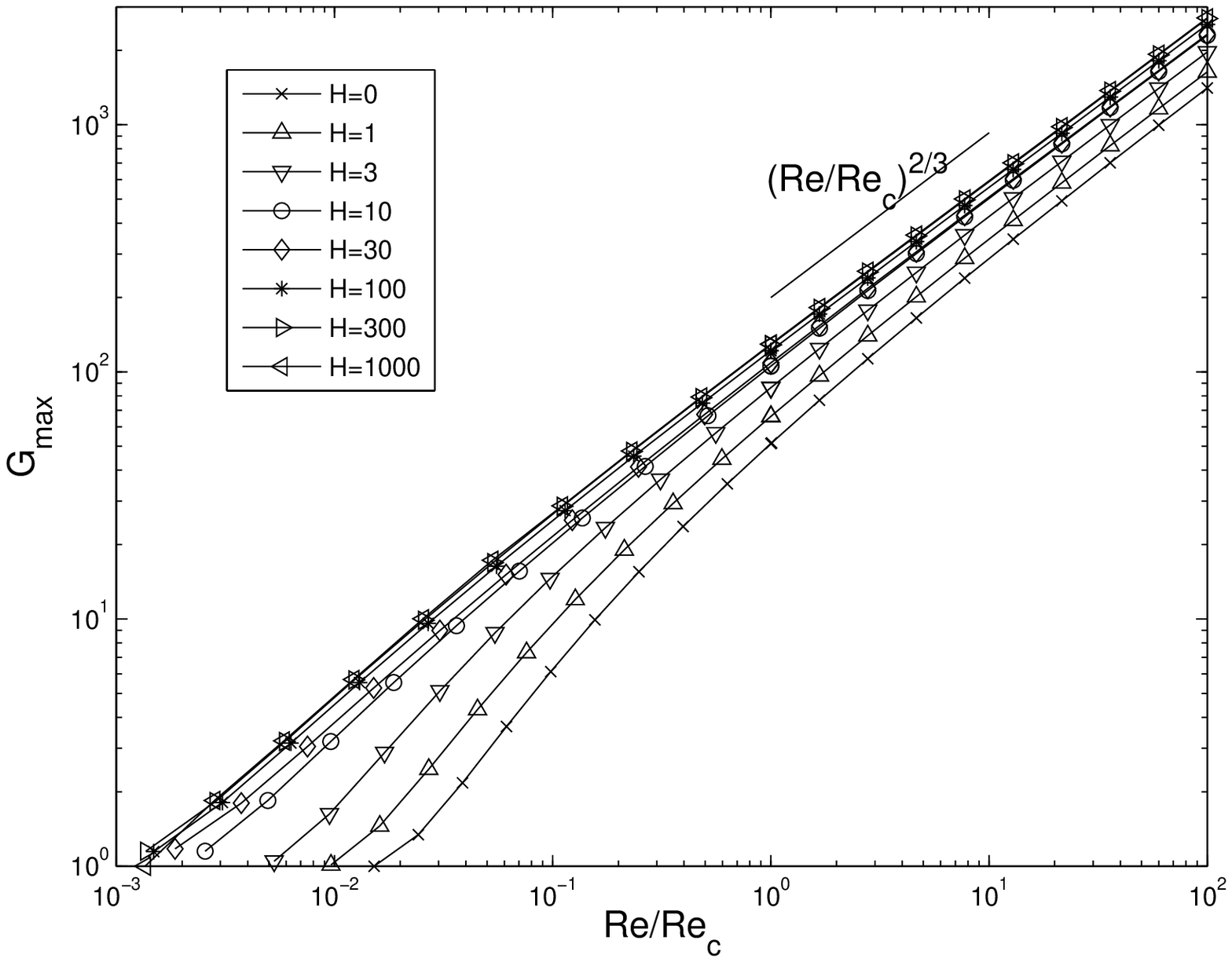}
\includegraphics[width=8.5cm]{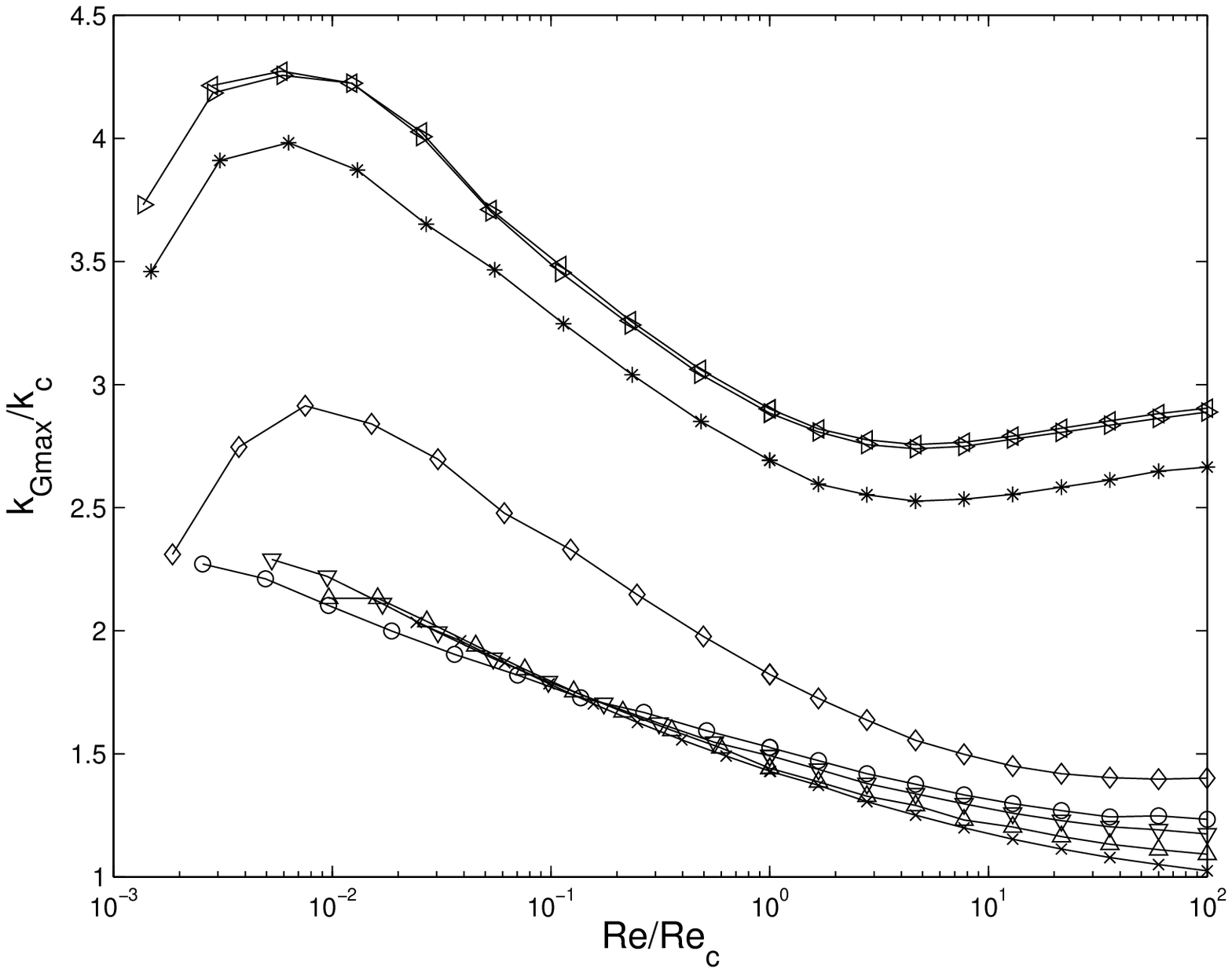}
\includegraphics[width=8.5cm]{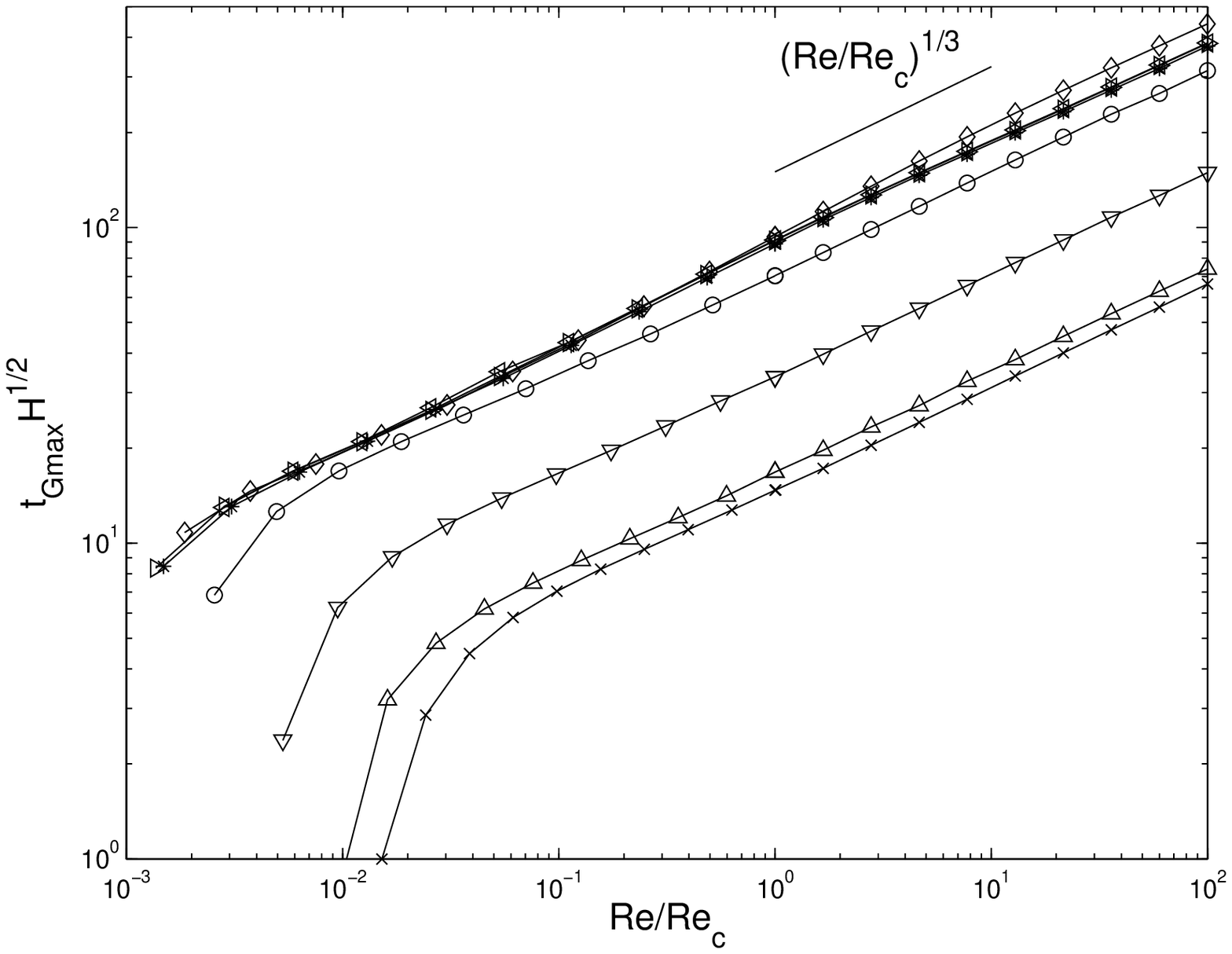}
\caption{Maximum transient growth $G_{max}$ over time and streamwise wavenumber  (top), streamwise wavenumber of maximum transient growth $k_{Gmax}$ (middle) 
 and time of maximum transient growth $t_{Gmax}$ (bottom) as a function of 
$Re/Re_c$ for $H\in\{0, 1, 3, 10, 30, 100, 300, 1000\}$. The curves for H=300 
and H=1000 cannot be distinguished, which proves once again that good 
asymptotics are reached for a value of $H$ of a few hundred. For $H=0$, 
$t_{Gmax}$ and not $t_{Gmax}H^{1/2}$ vs. $Re/Re_c$ is represented.} 
\label{fig:gmax}
\end{figure}

We shall now characterise more quantitatively how the optimal modes, their 
associated wavelength, growth rate and time of maximum growth rate vary with 
$H$ and $Re$. Figure \ref{fig:gmax_k} shows the variations of the local 
maximum $G_{mk}$ of 
$G(t)$ as well as those of $t_G$ as a function of $k$, for $H=100$ and $Re$ 
varying 
from the critical $Re_E(H=100)$ to $100Re_c(H=100)$. When transient 
growth exists, $G_{mk}(k)$ reaches a maximum $G_{max}$ for $k=k_m(Re,H)$.
As opposed to the linear stability, which only 
yields narrow bandwidths of unstable wavenumbers, some significant transient 
growth occurs on a broad, high pitched bandwidth extending much further than 
our maximum calculated value of $k/k_c=6$, for $H>100$.\\
Figure \ref{fig:gmax} shows the variations of $G_{max}$ as a
 function of $Re/Re_c(H)$ for different values of $H$. All curves are rather 
close to each other which means that the Maximum gain is mostly controlled
 by the parameter $Re/Re_c(H)$. This becomes exact as $H$ tends to infinity 
since all curves rapidly approach an asymptotic exponential law of the form
 $G_{max}\sim (Re/Re_c)^{2/3}$. 
This exponent of $Re$ is smaller than that of $2$ obtained for the maximum 
transient growth 
in 3D Poiseuille flow, mostly because the optimal perturbations for the 3D 
Poiseuille flow are 
 streamwise independent. Since such perturbation cannot exist in 
two-dimensional dynamics, the maximum transient growth is achieved by streamwise
dependent perturbations with a lower gain. The mechanism of their 
amplification is also different to that of the 3D ones, which results from 
the coupling between Orr-Sommerfeld and Squire modes. Squire modes are not 
present in 2D so transient growth results exclusively from the combination
of Orr-Sommerfeld modes which offers fewer possible combinations and therefore 
achieves a lower maximum gain. In spite of this, some significant transient 
growth 
(up to two orders of magnitude) occurs for $Re/Re_c$ below unity. 
$G_{max}$ is even between $7$ (for $Re/Re_c=10^{-2}$) and $2.5$ (for $Re/Re_c=1$) 
times greater than for the 2D Poiseuille flow ($H=0$) studied by \cite{reddy93}.
It is therefore reasonable to think that those optimal perturbations 
destabilise the flow
at Reynolds numbers significantly smaller than $Re_c$, and even at lower 
$Re/Re_c$ than in the case $H=0$. These optimal modes are even more likely to be 
present in supercritical flows as $G_{max}$ continues to increase as 
$(Re/Re_c)^{2/3}$ even for $Re/Re_c>1$, yielding gains of the order of $10^3$.\\
The wavenumber of maximum transient growth $k_{Gmax}$, and related time 
$t_{Gmax}$ as a function of 
$Re/Re_c$ are reported on figure \ref{fig:gmax} for different values of $H$. As 
for the gain $G_{max}$, the curves are quite close to each other and approach
 an asymptotic curve for high $H$. The time at which 
the optimal mode reaches maximal amplification $t_{Gmax}$ quickly approaches a 
power law 
 of the form  $t_{Gmax}H^{1/2}\sim(Re/Re_c)^{1/3}$ as $Re/Re_c$ increases, for 
all values of $H$ except $H=0$ for which $t_{Gmax}\sim(Re/Re_c)^{1/3}$. When 
$H$ tends to infinity, all curves merge into the same asymptotic one.
%
\section{Conclusion}
We have investigated the quasi-2D stability of an MHD flow in a rectangular 
duct using the 2D model equation derived by \cite{sm82} for flows in a 
transverse magnetic field. The terminology "quasi-2D" indicates here that 
although the whole study has been conducted using only two space variables, 
the three-dimensionality due to the presence of Hartmann layers is taken into 
account in the model through a linear friction term. This model equation
provides a simple 2D model for the transition to turbulence in a duct, the 
dynamics of which can be studied over a wide range of parameters at little 
computational expense. The price to pay for this simplicity is that all 
perturbations are assumed to have a Hartmann profile along the direction of 
the magnetic field. The weak three-dimensionality in the base profile as 
well as 3D perturbations are therefore neglected so the focus is on the quasi-2D
 dynamics of those layers.\\
  The critical modes for the linear stability analysis have been found to be 
antisymmetric Tollmien-Schlichting waves. 
In the limit $H\rightarrow \infty$ (in practise $H \gtrsim 200$), their linear 
stability is governed by the Reynolds number scaled on the thickness of the 
Shercliff layer $Re/H^{1/2}$ with a critical value of $48350$ providing a 
sufficient condition for stability (2D and 3D), and 
corresponding critical wavenumber $k_c/H^{1/2}=0.161532$.
For $H\gtrsim42$, a second type of unstable mode appears, made of 3 rows of 
vortices, symmetric about the duct axis. 
These parameters  
also govern the energy stability, which provides a sufficient condition for 
2D stability,  with critical parameters $Re_E/H^{1/2}=65.3288$ 
and $k_E/H^{1/2}=0.86348422$. For values of $Re$ between those two stability 
bounds, quasi-two dimensional perturbations undergo some significant transient
growth.  Here again, an asymptotic regime is reached for $H$ of a few hundred. 
For given $H$ and $Re$, and in the limit $H\rightarrow\infty$,  the 
maximum gain varies as $G_{max}\sim (Re/R_c)^{2/3}$, and it is achieved at time 
 $t_{Gmax}\sim H^{-1/2}(Re/R_c)^{1/3}$ by perturbations of wavelength of the 
order of $k_c$. This transient growth mechanism 
results from combinations of Orr-Sommerfeld modes and differs from that 
happening in 3D Poiseuille flows where the more efficient interaction between 
Squire and Orr-Sommerfeld modes leads to a more significant transient growth, 
with $G_{max}\sim Re^2$.
In subcritical regime, $G_{max}$ however reaches values up to 100 and 
continues to increase in supercritical regime. This 
indicates that quasi-2D perturbations most certainly destabilise the flow for 
values of $Re$ much smaller than $Re_c$, but also that even in supercritical 
regime, transient growth is more likely to drive the instability than 
normal mode exponential growth. The next step is then to perform some DNS of 
(\ref{eq:sm82}) in order to describe the non-linear 
evolution of these optimal perturbations, and find out the critical 
 value of $Re$ at which they destabilise the base flow.\\

Now the next question is that of the implications of these results for the 3D 
the stability of the real 3D flow. This shall be our last point. The 
experiments of \cite{moresco04} have shown that the Hartmann layer in a 
duct such as that from figure \ref{fig:3dconfig} with $a/L=1$  (so $H=Ha$) 
becomes unstable for $Re \geq Re^{(H)}=380 H$. We have shown that in the same 
configuration, Shercliff 
layers are unstable to quasi to two-dimensional perturbations for 
$Re\geq Re^{(S)} = 4.83 .10^4 H^{1/2}$. For any $H>H_c=16189$, $Re^{(S)}
<Re^{(H)}$. In other words, for $H>H_c$, the first instability to appear in 
the duct flow is that of the Shercliff layers, and not that of the Hartmann 
layers. The fact that quasi-2D perturbations undergo some significant transient 
growth in Shercliff layers for $Re$ much lower than $Re^{(S)}$ makes it likely 
that Shercliff layers become unstable to these perturbations at much lower 
Reynolds numbers. This, in turn, implies that the values $H_c$ of $H$ at which 
$Re^{(S)}<Re^{(H)}$ is much lower than $16189$.\\
The author doesn't however believe this casts any doubt on the results 
obtained by \cite{moresco04} which have been closely recovered by the numerical simulations of \cite{krasnov04}. It is quite possible that the transition they
identify is not one between a laminar duct flow and a flow with turbulent 
Hartmann layers but rather one between a quasi-2D turbulent flow with a stable 
Hartmann layer and a flow with turbulent Hartmann layers. Nevertheless, since 
the 
transition was detected by measuring the friction, one can be most assured that 
it is indeed a transition to turbulence in the Hartmann layer. The reason for 
this is that 2D turbulence involves mostly big vortices inducing a total 
friction which is hardly higher than that of the laminar state, and therefore 
very difficult to detect in their experiment. On the contrary, once turbulent, 
the Hartmann layer introduces strong three-dimensional perturbations 
responsible for a much higher friction.\\
This reasoning is of course only valid if the two-dimensional perturbations
analysed in this work are effectively responsible for the destabilisation of 
the real 3D duct flow. Whether 3D perturbations not taken into account here 
play the leading role or whether the three-dimensionality of the 
base profile affects the growth of quasi-2D perturbations 
are to this day open questions and shall be answered by full 3D numerical 
simulations. In this regard, some recent work suggests that perturbations which are 
invariant along the magnetic field lines, similar to the optimal perturbations 
we find in this work, may well be the ones driving the instability as soon as 
$Ha$ exceeds about $100$. This question should now be investigated in the 
duct configuration...
%


\end{document}